\documentclass[12pt]{article}

\usepackage{amssymb}
\usepackage[mathscr]{euscript}
\usepackage{graphicx}
\usepackage[all]{xypic}

\newtheorem{prop}{Proposition}[section]

\newtheorem{defi}{Definition}[section]
\newtheorem{lemm}{Lemma}[section]
\newtheorem{theo}{Theorem}[section]
\newtheorem{coro}{Corollary}[section]

\newcommand{\bbox}{\normalsize {}%
        \nolinebreak \hfill $\blacksquare$ \medbreak \par}

\newcommand{\Ni}{\hbox{ {\vrule height .22cm}{\leaders\hrule\hskip.2cm} }}
\newcommand{\iN}{\hbox{ {\leaders\hrule\hskip.2cm}{\vrule height .22cm} }}

\newcommand{\R}[1][]{\ensuremath{{\mathbb{R}^{#1}} }}

\def\<{\langle} \def\>{\rangle}

\title{Covariant Hamiltonian formalism for the calculus of variations with several
variables: Lepage--Dedecker versus de Donder--Weyl}
\author{
Fr\'ed\'eric H\'ELEIN\footnote{helein@math.jussieu.fr} \\
Institut de Mathématiques de Jussieu, UMR 7586,\\
Université Denis Diderot--Paris 7, Site de Chevaleret,\\
16 rue Clisson 75013 Paris
(France)
\\ \\
Joseph KOUNEIHER\footnote{kouneiher@paris7.jussieu.fr}\\
LUTH, CNRS UMR 8102\\ Observatoire de Paris - section Meudon \\5 Place Jules Janssen \\92195 Meudon Cedex \\ Universit\'e Paris 7\\
}

\begin{document}
\maketitle
\begin{center}
{\bf Abstract}
\end{center}
The main purpose in the present paper is to build a Hamiltonian
theory for fields which is consistent with the principles of
relativity. For this we consider detailed geometric pictures of
Lepage theories in the spirit of Dedecker and try to stress out
the interplay between the Lepage-Dedecker (LP) description and the
(more usual) de Donder-Weyl (dDW) one. One of the main points is the fact that the Legendre
transform in the dDW approach is replaced by a Legendre
correspondence in the LP theory\footnote{ This correspondence behaves differently: ignoring the singularities
whenever the Lagrangian is degenerate.}.

\section{Introduction}
\subsection{Presentation}
Multisymplectic formalisms are finite dimensional descriptions of
variational problems with several variables (or field theories for
physicists) analogue to the well-known Hamiltonian theory of point
mechanics. For example consider on the set of maps
$u:\Bbb{R}^n\longrightarrow \Bbb{R}$ a Lagrangian action of the
type
\[
{\cal L}[u]=\int_{\Bbb{R}^n}L(x,u(x),\nabla u(x))dx^1\cdots dx^n.
\]
Then it is well-known that the maps which are critical points of
${\cal L}$ are characterized by the Euler--Lagrange equation
${\partial \over \partial x^\mu}\left({\partial L\over \partial
(\partial _\mu u)}\right) = {\partial L\over \partial u}$. By
analogy with the Hamiltonian theory we can do the change of
variables $p^\mu:= {\partial L\over \partial (\partial _\mu u)}$
and define the Hamiltonian function
\[
H(x,u,p):= p^\mu{\partial u\over \partial x^\mu} - L(x,u,\nabla
u),
\]
where here $\nabla u = \left({\partial u\over \partial
x^\mu}\right)$ is a function of $(x,u,p)$ defined implicitly by
$p^\mu:= {\partial L\over \partial (\partial _\mu u)}(x,u,\nabla
u)$. Then the Euler-Lagrange equation is equivalent to the
generalized Hamilton system of equations
\begin{equation}\label{0.h}
\left\{
\begin{array}{ccc}
\displaystyle {\partial u\over \partial x^{\mu}} & =
& \displaystyle {\partial H\over \partial p^{\mu}}(x,u,p)\\
\displaystyle \sum_{\mu}{\partial p^{\mu}\over \partial x^{\mu}} &
= & \displaystyle - {\partial H\over \partial u}(x,u,p).
\end{array}
\right.
\end{equation}
This simple observation is the basis of a theory discovered by T.
de Donder \cite{deDonder} and H. Weyl \cite{Weyl} independently in
1935. This theory can be formulated in a geometric setting, an
analogue of the symplectic geometry, which is governed by the
Poincar\'e--Cartan $n$-form $\theta:=  e \omega + p^{\mu} du\wedge
\omega_{\mu}$ (where $\omega:= dx^1\wedge \cdots  \wedge dx^n$ and
$\omega_{\mu}:= \partial _\mu\iN \omega$) and its differential
$\Omega:=d\theta$,
often called multisymplectic (or polysymplectic form).\\

\noindent Although similar to mechanics this theory shows up deep
differences. In particular there exist other theories which are
analogues of Hamilton's one as for instance the first historical
one, constructed by C. Carath\'eodory in 1929 \cite{Caratheodory}.
In fact, as realized by T. Lepage in 1936 \cite{Lepage}, there are
infinitely many theories, due to the fact that one could fix
arbitrary the value of some tensor in the Legendre transform (see
also \cite{Rund}, \cite{GiaquintaHildebrandt}). Much later on, in
1953, P. Dedecker \cite{Dedecker} built a geometrical framework in
which all Lepage theories are embedded. The present paper, which
is a continuation of \cite{HeleinKouneiher}, is devoted to the
study of the Lepage--Dedecker theory. We also want to compare this
formalism with the more popular de Donder--Weyl theory.\\

\begin{figure}[h]\label{fig-pseudo}
\begin{center}
\includegraphics[scale=0.5]{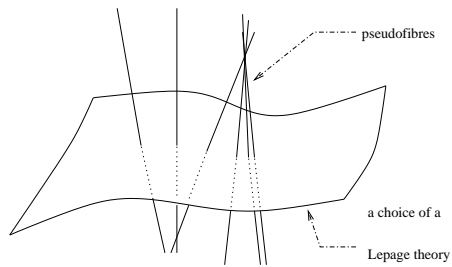}
\caption{\footnotesize Pseudofibers which intersect a submanifold
corresponding to the choice of a Lepage theory}
\end{center}
\end{figure}

\noindent First recall that the range of application of the de
Donder--Weyl theory is restricted in principle to variational
problems on sections of a bundle ${\cal F}$. The right framework
for it, as expounded e.g.\,in \cite{GIMMSY}, consists in using the
affine first jet bundle $J^1{\cal F}$ and its dual
$\left(J^1\right)^*{\cal F}$ as analogues of the tangent and the
cotangent bundles for mechanics respectively. For non degenerate
variational problems the Legendre transform induces a
diffeomorphism between $J^1{\cal F}$ and $\left(J^1\right)^*{\cal
F}$. In contrast the Lepage theories can be applied to more
general situations but involve, in general, many more variables
and so are more complicated to deal with, as noticed in
\cite{Kijowski2}. This is probably the reason why most papers on
the subject focus on the de Donder--Weyl theory,
e.g.\,\cite{Kanatchikov1}, \cite{GIMMSY}. The general idea of Dedecker
in \cite{Dedecker} for describing Lepage's theories is the
following: if we view variational problems as being defined on
$n$-dimensional submanifolds embedded in a $(n+k)$-dimensional
manifold ${\cal N}$, then what plays the role of the (projective)
tangent bundle to space-time in mechanics is the Grassmann bundle
$Gr^n{\cal N}$ of oriented $n$-dimensional subspaces of tangent
spaces to ${\cal N}$. The analogue of the cotangent bundle
in mechanics is $\Lambda^nT^*{\cal N}$. Note that
$\hbox{dim}Gr^n{\cal N} = n+k+nk$ so that
$\hbox{dim}\Lambda^nT^*{\cal N} = n+k+{(n+k)!\over n!k!}$ is
strictly larger than $\hbox{dim}Gr^n{\cal N} + 1$ unless $n=1$
(classical mechanics) or $k=1$ (submanifolds are hypersurfaces).
This difference between the dimensions reflects the multiplicity
of Lepage theories: as shown in \cite{Dedecker}, we substitute to
the Legendre transform a Legendre correspondence which associates to each
$n$-subspace $T\in Gr^n_q{\cal N}$ (a ``generalized velocity'')
 an affine subspace of $\Lambda^nT^*_q{\cal N}$ called
{\em pseudofibre} by Dedecker. Then two points in the same
pseudofiber do actually represent the same physical
(infinitesimal) state, so that the coordinates on
$\Lambda^nT^*{\cal N}$, called {\em momento\"\i des} by Dedecker
do not represent physically observable quantities. In this picture
any choice of a Lepage theory corresponds to a selection of a
submanifold of $\Lambda^nT^*{\cal N}$, which --- when the induced
Legendre transform is invertible --- intersects transversally each
pseudofiber at one point (see Figure \ref{fig-pseudo}): so the
Legendre correspondence specializes to a Legendre transform. For
instance the de Donder--Weyl theory can be recovered in this
setting by the restriction to some submanifold of
$\Lambda^nT^*{\cal N}$ (see Section 2.2).

\noindent In \cite{HeleinKouneiher} and in the present paper we
consider a geometric pictures of Lepage theories in the spirit of
Dedecker and we try to stress out the interplay between the
Lepage--Dedecker description and the de Donder--Weyl one. Roughly
speaking a comparison between these two points of view shows up
some analogy with some aspects of the projective geometry, for
which there is no perfect system of coordinates, but basically
two: the homogeneous ones, more symmetric but redundant (analogue
to the Dedecker description) and the local ones (analogue to the
choice of a particular Lepage theory like e.g.\,the de
Donder--Weyl one). Note that both points of view are based on the
same geometrical framework, a multisymplectic manifold:
\begin{defi}\label{0.def1}
Let ${\cal M}$ be a differential manifold. Let $n\in \Bbb{N}$ be
some positive integer. A smooth $(n+1)$-form $\Omega$ on ${\cal
M}$ is a {\bf multisymplectic} form if and only if
\begin{enumerate}
\item [(i)] $\Omega$ is non degenerate, i.e.\,$\forall m\in {\cal
M}$, $\forall \xi \in T_m{\cal M}$, if $\xi \iN \Omega_m = 0$,
then $\xi = 0$ \item [(ii)] $\Omega$ is closed, i.e.\,$d\Omega =
0$.
\end{enumerate}
Any manifold ${\cal M}$ equipped with a multisymplectic form
$\Omega$ will be called a {\bf multisymplectic} manifold.
\end{defi}
For the de Donder--Weyl theory ${\cal M}$ is
$\left(J^1\right)^*{\cal F}$ and for the Lepage--Dedecker theory
${\cal M}$ is $\Lambda^nT^*{\cal N}$. In both descriptions
solutions of the variational problem correspond to $n$-dimensional
submanifolds $\Gamma$ (analogues of Hamiltonian trajectories: we
call them {\bf Hamiltonian $n$-curves}) and are characterized by
the Hamilton equation $X\iN \Omega = (-1)^nd{\cal H}$, where $X$
is a $n$-multivector tangent to $\Gamma$, ${\cal H}$ is a
(Hamiltonian) function defined on ${\cal M}$ and by ``$\iN$'' we
mean the interior product.\\

\noindent In Section 2 we present a complete derivation of the
(Dedecker) Legendre correspondence and of the generalized Hamilton
equations. We use a method that does not rely on any
trivialization or connection on the Grassmannian bundle. A
remarkable property, which is illustrated in this paper through
the examples given in Paragraph 2.2.2, is that when $n$ and $k$ are
greater than 2, the Legendre correspondence is generically never
degenerate. The more spectacular example is when the Lagrangian
density is a constant function --- the most degenerate situation
one can think about --- then the Legendre correspondence is
well-defined almost everywhere except precisely along the de
Donder--Weyl submanifold. We believe that such a phenomenon was
not noticed before; it however may be useful when one deals for
example with the bosonic string theory with a skewsymmetric 2-form
on the target manifold (a ``$B$-field'', as discussed in
\cite{HeleinKouneiher} and in subsection 2.2, example 5) or with the Yang--Mills action in 4
dimensions with a topological term in the Lagrangian: then the de
Donder--Weyl formalism may fail but one can cure this
degenerateness by using another Lepage theory or by working in the
full Dedecker setting.\\

\noindent In this paper we also stress out another aspect of the
(Dedecker) Legendre correspondence: one expects that the resulting
Hamiltonian function on $\Lambda^nT^*{\cal N}$ should satisfy some
condition expressing the ``projective'' invariance along each
pseudofiber. This is indeed the case. On the one hand we observe
in Section 2.1 that any smoothly continuous deformation of a
Hamiltonian $n$-curve along directions tangent to the pseudofibers
remains a Hamiltonian $n$-curve\footnote{A property quite similar to a gauge theory behavior although of different meaning. Here we are interested by desingularizing the theory and avoid the problems related to the presence of a constraints.} (Corollary \ref{2.2.3.coro1}). On
the other hand we give in Section 4.3 an intrinsic
characterization of the subspaces tangent to pseudofibers. This
motivates the definition given in Section 3.3 of the {\em
generalized pseudofiber directions} on any multisymplectic
manifold.\\

\noindent Beside these properties in this paper
and in its companion paper \cite{HK1b} we wish to address other kind of questions
related to the physical gain of these theories: the main
advantage of multisymplectic formalisms is to offer us a
Hamiltonian theory which is consistent with the principles of
Relativity, i.e.\,being {\em covariant}. Recall for instance that
for all the multisymplectic formalisms which have been proposed
one does not need to use a privilege time coordinate. One of our
ambitions in this paper was to try to extend this democracy
between space and time coordinates to the coordinates on fiber
manifolds (i.e.\,along the fields themselves). This is quite in
the spirit of the Kaluza--Klein theory and its modern avatars:
11-dimensional supergravity, string theory and M-theory. This
concern leads us naturally to replace de Donder--Weyl by the
Dedecker theory. In particular we do not need in our formalism to
split the variables into the horizontal (i.e.\,corresponding to
space-time coordinates) and vertical (i.e.\,non horizontal)
categories.\\

\noindent Moreover we may think that we start from a
(hypothetical) geometrical model where space-time and fields
variables would not be distinguished {\em a priori} and then ask
how to make sense of a space-time coordinate function  (that we
call a ``$r$-regular'' in Section 3.2) ? A variant of this
question would be how to define a constant time hypersurface (that
we call a ``slice'' in Section 3.2) without referring to a given
space-time background ? We propose in Section 3.2 a definition of
$r$-regular functions and of slices which, roughly speaking,
requires a slice to be transversal to all Hamiltonian $n$-curves.
Here the idea is that the dynamics only (i.e.\,the
Hamiltonian equation) should determine what are the slices. We
give in Section 4.2 a characterization of these slices in the case
where the multisymplectic manifold is $\Lambda^nT^*{\cal N}$.\\

\noindent These questions are connected to the concept of
observable functionals over the set of solutions of the Hamilton
equation. First because by using a codimension $r$ slice $\Sigma$
and an $(n-r)$-form $F$ on the multisymplectic manifold one can
define such a functional by integrating $F$ over the the
intersection of $\Sigma$ with a Hamiltonian curve. And second
because one is then led to impose conditions on $F$ in such a way
that the resulting functional carries only dynamical information.
The analysis of these conditions is the subject of our companion
paper \cite{HK1b}. And we believe that the conditions required on
these forms are connected with the definitions of
$r$-regular functions given in this paper, although we have not completely
elucidated this point.\\

\noindent Lastly in a future paper \cite{HeleinKouneiher1.1} we
investigate gauge theories, addressing the question of how to
formulate a fully covariant multisymplectic for them. Note that
the Lepage--Dedecker theory expounded here does not answer this
question completely, because a connection cannot be seen as a
submanifold. We will show there that it is possible to adapt this
theory and that a convenient covariant framework consists in
looking at gauge fields as {\em equivariant} submanifolds over the
principal bundle of the theory, i.e.\,satisfying some suitable
zeroth and first order differential constraints.

\subsection{Notations}

\noindent The Kronecker symbol $\delta^{\mu}_{\nu}$ is equal to 1 if
$\mu = \nu$ and equal to 0 otherwise. We shall also set
\[
\delta^{\mu_1\cdots \mu_p}_{\nu_1\cdots \nu_p}:=
\left| \begin{array}{ccc}
\delta^{\mu_1}_{\nu_1} & \dots  & \delta^{\mu_1}_{\nu_p}\\
\vdots & & \vdots \\
\delta^{\mu_p}_{\nu_1} & \dots & \delta^{\mu_p}_{\nu_p}
\end{array}\right| .
\]
In most examples, $\eta_{\mu\nu}$ is a constant
metric tensor on $\Bbb{R}^n$ (which may be Euclidean or Minkowskian).
The metric on his dual space his $\eta^{\mu\nu}$. Also, $\omega$ will
often denote a volume form on some space-time: in local coordinates
$\omega=dx^1\wedge \cdots \wedge dx^n$ and we will use several times the notation
$\omega_\mu:= {\partial \over \partial x^\mu}\iN \omega$,
$\omega_{\mu\nu}:= {\partial \over \partial x^\mu}\wedge
{\partial \over \partial x^\nu}\iN \omega$, etc. Partial derivatives
${\partial \over \partial x^\mu}$ and ${\partial \over \partial p_{\alpha_1\cdots \alpha_n}}$
will be sometime abbreviated by $\partial _\mu$ and $\partial ^{\alpha_1\cdots \alpha_n}$
respectively.\\

\noindent
When an index or a symbol is omitted in the middle of a
sequence of indices or symbols, we denote this omission by $\widehat{\,}$.
For example $a_{i_1\cdots \widehat{i_p}\cdots i_n}:=
a_{i_1\cdots i_{p-1}i_{p+1}\cdots i_n}$,
$dx^{\alpha_1}\wedge \cdots \wedge \widehat{dx^{\alpha_\mu}}\wedge \cdots \wedge dx^{\alpha_n} :=
dx^{\alpha_1}\wedge \cdots \wedge dx^{\alpha_{\mu-1}}\wedge dx^{\alpha_{\mu+1}}\wedge \cdots \wedge dx^{\alpha_n}$.\\

\noindent If ${\cal N}$ is a manifold
and ${\cal FN}$ a fiber bundle over
${\cal N}$, we denote by $\Gamma({\cal N},{\cal FN})$ the set of smooth sections of
${\cal FN}$. Lastly we use the following notations concerning the exterior algebra of multivectors and
differential forms.
If ${\cal N}$ is a differential $N$-dimensional manifold and $0\leq k\leq N$,
$\Lambda^kT{\cal N}$ is the bundle over ${\cal N}$ of $k$-multivectors
($k$-vectors in short)
and $\Lambda^kT^{\star}{\cal N}$ is the bundle of differential forms of degree $k$
($k$-forms in short). Setting $\Lambda T{\cal N}:= \oplus_{k=0}^N\Lambda^kT{\cal N}$
and $\Lambda T^{\star}{\cal N}:= \oplus_{k=0}^N\Lambda^kT^{\star}{\cal N}$, there exists
a unique duality evaluation map between $\Lambda T{\cal N}$ and $\Lambda T^{\star}{\cal N}$
such that for every decomposable $k$-vector field $X$, i.e.\,of the form
$X=X_1\wedge \cdots \wedge X_k$, and for every $l$-form $\mu$, then
$\langle X,\mu\rangle = \mu(X_1,\cdots ,X_k)$ if $k=l$ and $=0$ otherwise.
Then interior products $\iN $ and $\Ni$ are operations
defined as follows. If $k\leq l$, the product
$\iN :\Gamma({\cal N},\Lambda^kT{\cal N})\times \Gamma({\cal N},\Lambda^lT^{\star}{\cal N})\longrightarrow
\Gamma({\cal N},\Lambda^{l-k}T^{\star}{\cal N})$
is given by
$$\langle Y,X\iN \mu\rangle = \langle X\wedge Y,\mu\rangle ,\quad
\forall (l-k)\hbox{-vector }Y.$$
And if $k\geq l$, the product $\Ni :\Gamma({\cal N},\Lambda^kT{\cal N})\times
\Gamma({\cal N},\Lambda^lT^{\star}{\cal N})\longrightarrow
\Gamma({\cal N},\Lambda^{k-l}T{\cal N})$ is given by
\[
\langle X\Ni \mu,\nu\rangle = \langle X,\mu\wedge \nu\rangle,\quad
\forall (k-l)\hbox{-form }\nu.
\]

\section{The Lepage--Dedecker theory}
We expound here a Hamiltonian formulation of a large class of second order variational problems in
an intrinsic way. Details and computations in coordinates can be
found in \cite{Kanatchikov1}, \cite{HeleinKouneiher}.

\subsection{Hamiltonian formulation of variational problems with several variables}
\subsubsection{Lagrangian formulation}
The category of Lagrangian variational problems we start with is described as
follows. We consider $n,k\in \Bbb{N}^*$ and a smooth manifold ${\cal N}$ of dimension
$n+k$; ${\cal N}$ will be equipped with a closed nowhere vanishing ``space-time volume''
$n$-form $\omega$. We define
\begin{itemize}
\item the Grassmannian bundle $Gr^n{\cal N}$, it is the fiber bundle over
${\cal N}$ whose fiber over $q\in{\cal N}$ is $Gr^n_q{\cal N}$,
the set of all oriented $n$-dimensional vector subspaces of $T_q{\cal N}$.
\item the subbundle $Gr^\omega{\cal N}:= \{(q,T)\in Gr^n{\cal N}/
\omega_{q|T}>0\}$.
\item the set ${\cal G}^\omega$, it is the set of all oriented $n$-dimensional submanifolds
$G\subset {\cal N}$, such that $\forall q\in G$,
$T_qG\in Gr^\omega_q{\cal N}$ (i.e.\,the restriction of $\omega$ on
$G$ is positive everywhere).
\end{itemize}
Lastly we consider any Lagrangian density $L$, i.e.\,a smooth function
$L:Gr^\omega{\cal N}\longmapsto \Bbb{R}$. Then the Lagrangian of any $G\in {\cal G}^\omega$
is the integral
\begin{equation}\label{2.2.1.action}
{\cal L}[G]:= \int_GL\left(q,T_qG\right)\omega
\end{equation}
We say that a submanifold $G\in {\cal G}^\omega$ is a {\bf critical point of ${\cal L}$}
if and only if, for any compact $K\subset {\cal N}$, $G\cap K$ is a critical
point of ${\cal L}_K[G]:= \int_{G\cap K}L\left(q,T_qG\right)\omega$ with respect to
variations with support in $K$.\\

\noindent It will be useful to represent $Gr^n{\cal N}$
differently, by means of $n$-vectors. For any $q\in {\cal N}$, we
define $D^n_q{\cal N}$ to be the set of decomposable
$n$-vectors\footnote{another notation for this set would be
$D\Lambda^nT_q{\cal N}$, for it reminds that it is a subset of
$\Lambda^nT_q{\cal N}$, but we have chosen to lighten the
notation.}, i.e.\,elements $z\in \Lambda^nT_q{\cal N}$ such that
there exists $n$ vectors $z_1$,...,$z_n\in T_q{\cal N}$ satisfying
$z = z_1\wedge \cdots \wedge z_n$. Then $D^n{\cal N}$ is the fiber
bundle whose fiber at each $q\in {\cal N}$ is $D^n_q{\cal N}$.
Moreover the map
$$\begin{array}{ccc}
D^n_q{\cal N} & \longrightarrow & Gr^n_q{\cal N}\\
z_1\wedge \cdots \wedge z_n & \longmapsto & T(z_1,\cdots ,z_n),
\end{array}$$
where $T(z_1,\cdots ,z_n)$ is the vector space spanned and oriented by $(z_1,\cdots ,z_n)$,
induces a diffeomorphism between $\left(D^n_q{\cal N}\setminus \{0\}\right)/\Bbb{R}^*_+$
and $Gr^n_q{\cal N}$. If we set also
$D^\omega_q{\cal N}:=\{(q,z)\in D^n_q{\cal N}/ \omega_q(z)=1\}$, the same map allow
² us also to identify $Gr^\omega_q{\cal N}$ with $D^\omega_q{\cal N}$.\\

\noindent This framework includes a large variety of situations as illustrated below.\\

\noindent {\bf Example 1} --- {\em Classical point mechanics} --- The motion of a point moving
in a manifold ${\cal Y}$ can be represented by its graph
$G\subset {\cal N}:=\Bbb{R}\times {\cal Y}$. If $\pi:{\cal N}\longrightarrow \Bbb{R}$
is the canonical projection and $t$ is the time coordinate on $\Bbb{R}$, then $\omega:=
\pi^*dt$.\\
\noindent {\bf Example 2} --- {\em Maps between manifolds} --- We
consider maps $u:{\cal X}\longrightarrow {\cal Y}$, where ${\cal
X}$ and ${\cal Y}$ are manifolds of dimension $n$ and $k$
respectively and ${\cal X}$ is equipped with some non vanishing
volume form $\omega$. A first order Lagrangian density can
represented as a function $l: T{\cal Y}\otimes _{{\cal X}\times
{\cal Y}}T^{\star}{\cal X}\longmapsto \Bbb{R}$, where $T{\cal
Y}\otimes _{{\cal X}\times {\cal Y}}T^{\star}{\cal X}:= \{
(x,y,v)/(x,y)\in {\cal X}\times {\cal Y},v\in T_y{\cal Y}\otimes
T_x^*{\cal X}\}$. (We use here a notation which exploits the
canonical identification of $T_y{\cal Y}\otimes T_x^*{\cal X}$
with the set of linear mappings from $T_x{\cal X}$ to $T_y{\cal
Y}$). The action of a map $u$ is
$$\ell [u]:= \int_{\cal X}l(x,u(x),du(x))\omega.$$
In local coordinates $x^{\mu}$ such that $\omega=dx^1\wedge \cdots \wedge dx^n$,
critical points of $\ell$
satisfy the Euler-Lagrange equation
$\sum_{\mu=1}^n{\partial \over \partial x^{\mu}}\left(
{\partial l \over \partial v^i_{\mu}}(x,u(x),du(x))\right) =
{\partial l \over \partial y^i}(x,u(x),du(x))$, $\forall i=1,\cdots ,k$.\\
Then we set ${\cal N}:={\cal X}\times {\cal Y}$ and denoting by $\pi:{\cal N}\longrightarrow
{\cal X}$ the canonical projection, we use the volume form $\omega\simeq \pi^*\omega$.
Any map $u$ can be represented by its graph
$G_u:=\{(x,u(x))/\,x\in {\cal X}\}\in {\cal G}^\omega$, (and conversely if
$G\in {\cal G}^\omega$ then the condition $\omega_{|G}>0$ forces $G$ to
be the graph of some map). For all $(x,y)\in{\cal N}$ we also have a diffeomorphism
$$\begin{array}{ccc}
T_y{\cal Y}\otimes T_x^*{\cal X} & \longrightarrow & Gr^\omega_{(x,y)}{\cal N}
\simeq D^\omega_{(x,y)}{\cal N}\\
v & \longmapsto & T(v),
\end{array}$$
where $T(v)$ is the graph of the linear map $v:T_x{\cal X}\longrightarrow T_y{\cal Y}$.
Then if we set $L(x,y,T(v)):= l(x,y,v)$, the action defined by (\ref{2.2.1.action})
coincides with $\ell$.\\
\noindent {\bf Example 3} --- {\em Sections of a fiber bundle} --- This is a particular
case of our setting, where ${\cal N}$ is the total space of a fiber bundle with base
manifold ${\cal X}$. The set ${\cal G}^\omega$ is then just the set of smooth sections.\\

\subsubsection{The Legendre correspondence}
\noindent Now we consider the manifold $\Lambda^nT^*{\cal N}$ and
the projection mapping $\Pi:\Lambda^nT^*{\cal N}\longrightarrow
{\cal N}$. We shall denote by $p$ an $n$-form in the fiber
$\Lambda^nT^*_q{\cal N}$. There is a canonical $n$-form $\theta$
called the {\em Poincar\'e--Cartan} form defined on
$\Lambda^nT^*{\cal N}$ as follows: $\forall (q,p)\in
\Lambda^nT^*{\cal N}$, $\forall X_1,\cdots ,X_n\in
T_{(q,p)}\left(\Lambda^nT^*{\cal N}\right)$,

\[
\theta_{(q,p)}(X_1,\cdots ,X_n):= p\left( \Pi_*X_1,\cdots ,\Pi_*X_n\right)
 = \langle \Pi_*X_1\wedge \cdots \wedge \Pi_*n,p\rangle ,
\]
where $\Pi_*X_\mu := d\Pi_{(q,p)}(X_\mu)$.
If we use local coordinates $\left(q^\alpha\right)_{1\leq \alpha\leq n+k}$ on
${\cal N}$, then a basis of $\Lambda^nT^*_q{\cal N}$ is the family
$\left( dq^{\alpha_1}\wedge \cdots \wedge dq^{\alpha_n}\right)_{1\leq \alpha_1<
\cdots <\alpha_n \leq n+k}$ and we denote by $p_{\alpha_1\cdots \alpha_n}$ the coordinates
on $\Lambda^nT^*_q{\cal N}$ in this basis. Then $\theta$ writes
\begin{equation}\label{2.2.2.theta}
\theta:= \sum_{1\leq \alpha_1<\cdots <\alpha_n\leq n+k}
p_{\alpha_1\cdots \alpha_n}dq^{\alpha_1}\wedge \cdots \wedge dq^{\alpha_n}.
\end{equation}
Its differential is the {\bf multisymplectic} form $\Omega:=
d\theta$ and
will play the role of generalized symplectic form.\\

\noindent In order to build the analogue of the Legendre transform we consider the fiber bundle
$Gr^\omega{\cal N}\times _{\cal N}\Lambda^nT^*{\cal N}:=\{(q,z,p)/q\in {\cal N},
z\in Gr^\omega_q{\cal N}\simeq D^\omega_q{\cal N},
p\in \Lambda^nT^*_q{\cal N}\}$ and we denote by
$\widehat{\Pi}:Gr^\omega{\cal N}\times _{\cal N}\Lambda^nT^*{\cal N}
\longrightarrow {\cal N}$ the canonical projection. To summarize:
\begin{center}
\hskip15mm
\xymatrix{
& Gr^\omega{\cal N}\times _{\cal N}\Lambda^nT^*{\cal N} \ar@{->}[d]^{\Pi^L} \ar@{->}[r]^{\Pi^H} \ar@{->}[dr]^{\widehat{\Pi}} &
    \Lambda^nT^*{\cal N} \ar@{->}[d]^\Pi & {\cal M}\ar@{->}[l]_{\imath} \ar@{->}[dl]^{\Pi_{|{\cal M}}} \\
Gr^n{\cal N} & Gr^\omega{\cal N} \ar@{->}[l]^{\imath} \ar@{->}[r] &
    {\cal N} &
}
\end{center}

\noindent
We define on $Gr^\omega{\cal N}\times _{\cal N}\Lambda^nT^*{\cal N}$ the function
\[
W(q,z,p):= \langle z,p\rangle -L(q,z).
\]
Note that for each $(q,z,p)$ there a vertical subspace $V_{(q,z,p)}\subset
T_{(q,z,p)} ( Gr^\omega{\cal N}\times _{\cal N}\Lambda^nT^*{\cal N})$,
which is canonically defined as the kernel of
\[
d\widehat{\Pi}_{(q,z,p)}:T_{(q,z,p)}
\left( Gr^\omega{\cal N}\times _{\cal N}\Lambda^nT^*{\cal N}\right)
\longrightarrow T_q{\cal N}.
\]
We can further split $V_{(q,z,p)}\simeq T_zD^\omega_q{\cal N}
\oplus T_p\Lambda^nT^*_q{\cal N}$, where $T_zD^\omega_q{\cal N}\simeq
\hbox{Ker}d\Pi^H_{(q,z,p)}$ and $T_p\Lambda^nT^*_q{\cal N}\simeq
\hbox{Ker}d\Pi^L_{(q,z,p)}$.
Then, for any function $F$ defined on $Gr^\omega{\cal N}\times _{\cal N}\Lambda^nT^*{\cal N}$,
we denote respectively by $\partial F/\partial z(q,z,p)$
and $\partial F/\partial p(q,z,p)$ the restrictions of the differential\footnote{However
in order to make sense of ``$\partial F/\partial q(q,z,p)$'' we would need to define a ``horizontal''
subspace of $T_{(q,z,p)} \left( Gr^\omega{\cal N}\times _{\cal N}\Lambda^nT^*{\cal N}\right)$,
which requires for instance the use of a connection on the bundle
$Gr^\omega{\cal N}\times _{\cal N}\Lambda^nT^*{\cal N}\longrightarrow {\cal N}$. Indeed such a horizontal subspace prescribes a inertial law on $\cal{N}$, such a law would have a sense on a Galilee or Minkowski space-time but not in general relativity.}
$dF_{(q,z,p)}$ on respectively $T_zD^\omega_q{\cal N}$ and $T_p\Lambda^nT^*_q{\cal N}$.\\

\noindent
Instead of a Legendre transform we shall rather use a Legendre correspondence:
we write
\begin{equation}\label{2.2.2.corr}
(q,z) \longleftrightarrow (q,p)\quad \hbox{if and only if}
\quad {\partial W\over \partial z}(q,z,p) = 0.
\end{equation}

\noindent Let us try to picture geometrically the situation (see figure \ref{fig-tdn}):
\begin{figure}[h]\label{fig-tdn}
\begin{center}
\includegraphics[scale=0.5]{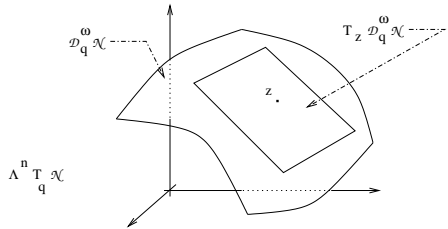}
\caption{\footnotesize $T_zD_q^\omega{\cal N}$ is a vector
subspace of $\Lambda^nT_q{\cal N}$}
\end{center}
\end{figure}
$D_q^\omega{\cal N}$
is a smooth submanifold of dimension $nk$ of the vector space $\Lambda^nT_q{\cal N}$,
which is of dimension ${(n+k)!\over n!k!}$; $T_zD_q^\omega{\cal N}$ is thus a
vector subspace of $\Lambda^nT_q{\cal N}$. And ${\partial L\over \partial z}(q,z)$
or ${\partial W\over \partial z}(q,z,p)$
can be understood as linear forms on $T_zD_q^\omega{\cal N}$ whereas
$p\in \Lambda^nT^*_q{\cal N}$ as a linear form on $\Lambda^nT_q{\cal N}$.
So the meaning of the right hand side of (\ref{2.2.2.corr}) is that
the restriction of $p$ at $T_zD_q^\omega{\cal N}$ coincides with
${\partial L\over \partial z}(q,z,p)$:
\begin{equation}\label{2.2.2.p=dl}
p_{|T_zD_q^\omega{\cal N}} = {\partial L\over \partial z}(q,z).
\end{equation}
Given $(q,z)\in Gr^\omega{\cal N}$ we define the {\bf enlarged
pseudofiber} in $q$ to be:
\[
P_q(z):=\{p\in \Lambda^nT^*_q{\cal N}/{\partial W\over \partial z}(q,z,p) = 0\}.
\]
In other words, $p\in P_q(z)$ if it is a solution of (\ref{2.2.2.p=dl}).
Obviously $P_q(z)$ is not empty; moreover given some $p_0\in P_q(z)$,
\begin{equation}\label{2.2.2.p-p}
p_1\in P_q(z),\ \Longleftrightarrow \ p_1-p_0\in
\left( T_zD_q^\omega{\cal N}\right)^{\perp}:=
\{p\in \Lambda^nT^*_q{\cal N}/\forall \zeta\in T_zD_q^\omega{\cal N},p(\zeta)=0\}.
\end{equation}
So $P_q(z)$ is an affine subspace of $\Lambda^nT^*_q{\cal N}$ of
dimension ${(n+k)!\over n!k!} - nk$. Note that in case where $n=1$
(the classical mechanics of point) then dim\,$P_q(z)=1$: this is
due to the fact that we are still free to fix arbitrarily the
momentum component dual to the time (i.e.\,the energy)\footnote{a
simple but more interesting example is provided by variational
problems on maps $u:\Bbb{R}^2\longrightarrow \Bbb{R}^2$. Then one
is led to the multisymplectic manifold
$\Lambda^2T^{\star}\Bbb{R}^4$. And given any $(q,z)\in
Gr^\omega\R^4$ the enlarged pseudofiber $P_q(z)\subset
\Lambda^2T^\star\R^4$ is an affine plane parallel to
$\Bbb{R}\left[ \left( v_1^1v_2^2 -v_1^2v_2^1 \right) dx^1\wedge
dx^2 - \epsilon_{ij}v_\nu^j dy^i\wedge dx^\nu + dy^1\wedge dy^2
\right] \oplus \Bbb{R}dx^1\wedge dx^2$, where (using the notations
of Example 2) $T(v)=z$. For details see Paragraph 2.2.2.}.\\

\noindent
We now define
\[
{\cal P}_q:= \bigcup_{z\in D^\omega_q{\cal N}}P_q(z) \subset \Lambda^nT^*_q{\cal N},
\quad \forall q\in {\cal N}
\]
and we denote by ${\cal P}:=\cup_{q\in {\cal N}}{\cal P}_q$
the associated bundle over ${\cal N}$. We also
let, for all $(q,p)\in \Lambda^nT^*{\cal N}$,
\[
Z_q(p):=\{ z\in Gr^\omega_q{\cal N}/ p\in P_q(z)\}.
\]
It is clear that $Z_q(p)\neq \emptyset \Longleftrightarrow p\in {\cal P}_q$.
Now in order to go further we need to choose some submanifold ${\cal M}_q\subset {\cal P}_q$,
its dimension is not fixed {\em a priori}.\\

\noindent {\bf Legendre Correspondence Hypothesis} --- {\em We
assume that there exists a subbundle manifold ${\cal M}\subset
{\cal P}\subset \Lambda^nT^*{\cal N}$ over ${\cal N}$ where $\dim
{\cal M}=:M$ such that,
\begin{itemize}
\item for all $q\in {\cal N}$ the fiber ${\cal M}_q$ is a smooth submanifold,
possibly with boundary, of dimension $1\leq M-n-k\leq {(n+k)!\over n!k!}$
\item for any $(q,p)\in {\cal M}$, $Z_q(p)$ is a non empty smooth connected
submanifold of $Gr^\omega_q{\cal N}$
\item if $z_0\in Z_q(p)$, then we have $Z_q(p)=\{z\in D^\omega_q{\cal N}/
\forall \dot{p}\in T_p{\cal M}_q, \langle z-z_0,\dot{p}\rangle = 0\}$.
\end{itemize}
}
\noindent
{\em Remark} --- In the case where $M={(n+k)!\over n!k!}+n+k$, then ${\cal M}_q$
is an open subset of $\Lambda^nT^*_q{\cal N}$ and so $T_p{\cal M}_q\simeq
\Lambda^nT^*_q{\cal N}$. Hence the last assumption of the Legendre Correspondence
Hypothesis means that $Z_q(p)$ is reduced to a point. In general this condition
will imply that the inverse correspondence can be rebuild by using the Hamiltonian
function (see Lemma \ref{2.2.2.dh} below).

\begin{lemm}\label{2.2.2.h=cste}
Assume that the Legendre correspondence hypothesis is true. Then
for all $(q,p)\in {\cal M}$, the restriction of $W$ to $\{q\}
\times Z_q(p)\times \{p\}$ is constant.
\end{lemm}
{\em Proof} --- Since $Z_q(p)$ is smooth and connected, it suffices to prove that
$W$ is constant along any smooth path inside $\{(q,z,p)/q,p\hbox{ fixed },z\in Z_q(p)\}$.
Let $s\longmapsto z(s)$ be a smooth path with values into $Z_q(p)$, then
\[
{d \over d s}\left( W(q,z(s),p)\right)  =
{\partial W\over \partial z}(q,z(s),p)
\left( {dz\over ds}\right) = 0,
\]
because of (\ref{2.2.2.corr}). \bbox

\noindent
A straightforward consequence of Lemma \ref{2.2.2.h=cste} is that we can define the
{\bf Hamiltonian function} ${\cal H}:{\cal M}\longrightarrow \Bbb{R}$ by
\[
{\cal H}(q,p):= W(q,z,p),\quad \hbox{where }z\in Z_q(p),\hbox{ i.e. }
{\partial W\over \partial z}(q,z,p) = 0.
\]

\noindent

Any function $f$ constructed this way will be called {\bf Legendre Image Hamiltonian function}.
In the following, for all $(q,p)\in {\cal M}$ and for all $z\in D^n_q{\cal N}$
we denote by
\[
\begin{array}{cccc}
z_{|T_p{\cal M}_q}: & T_p{\cal M}_q & \longrightarrow & \Bbb{R}\\
 & \dot{p} & \longmapsto & \langle z,\dot{p}\rangle
\end{array}
\]
the linear map induced by $z$ on $T_p{\cal M}_q$. Then:
\begin{lemm}\label{2.2.2.dh}
Assume that the Legendre Correspondence Hypothesis is true.
Then\footnote{The advised Reader may expect to have also the
relation ``${\partial {\cal H}\over \partial q}(q,p) = - {\partial
L\over \partial q}(q,z)$''. But as remarked above the meaning of
${\partial {\cal H}\over \partial q}$ and ${\partial L\over
\partial q}$ is not clearly defined, because we did not introduce
a connection on the bundle $Gr^\omega{\cal N}\times _{\cal
N}\Lambda^nT^*{\cal N}$. This does not matter and we shall make
the economy of this relation later ! (cf footnote 2)}
\begin{enumerate}
\item $\forall (q,p)\in {\cal M}$ and $\forall z\in Z_q(p)$,
\begin{equation}\label{2.2.2.hl}
{\partial {\cal H}\over \partial p}(q,p) = z_{|T_p{\cal M}_q}.
\end{equation}
As a corollary of the above formula, $z_{|T_p{\cal M}_q}$ does not
depend on the choice of $z\in Z_q(p)$. \item Conversely if
$(q,p)\in {\cal M}$ and $z\in D^\omega_q{\cal N}$ satisfy
condition (\ref{2.2.2.hl}), then $z\in Z_q(p)$ or equivalently
$p\in P_q(z)$.
\end{enumerate}
\end{lemm}

\noindent {\em Proof} --- Let $(q,p)\in {\cal M}$ and
$(0,\dot{p})\in T_{(q,p)}{\cal M}$, where $\dot{p}\in T_p{\cal
M}_q$. In order to compute $d{\cal H}_{(q,p)}(0,\dot{p})$, we
consider a smooth path $s\longmapsto (q,p(s))$ with values into
${\cal M}_q$ whose derivative at $s=0$ coincides with
$(0,\dot{p})$. We can further lift this path into another one
$s\longmapsto (q,z(s),p(s))$ with values into $Gr^\omega_q{\cal
N}\times {\cal M}_q$, in such a way that $z(s)\in Z_{q}(p(s))$,
$\forall s$. Then using (\ref{2.2.2.p=dl}) we obtain
\[
\begin{array}{ccl}
\displaystyle {d\over ds}\left({\cal H}(q,p(s))\right)_{|s=0} & = &
\displaystyle {d\over ds}\left( \langle z(s),p(s)\rangle - L(q,p(s))\; \right)_{|s=0}\\
 & = & \displaystyle
\langle \dot{z},p\rangle + \langle z,\dot{p}\rangle -
{\partial L\over \partial z}(q,z)(\dot{z}) =
\langle z,\dot{p}\rangle ,
\end{array}
\]
from which (\ref{2.2.2.hl}) follows. This proves (i).\\
The proof of (ii) uses the Legendre Correspondence Hypothesis:
consider $z,z_0\in D^n_q{\cal N}$ and assume that $z_0\in Z_q(p)$ and that
$z$ satisfies (\ref{2.2.2.hl}).
Then by applying the conclusion (i) of the Lemma to $z_0$ we deduce
that $\partial {\cal H}/\partial p(q,p) = z_{0|T_p{\cal M}_q}$ and thus
$(z-z_0)_{|T_p{\cal M}_q}=0$. Hence by the Legendre Correspondence Hypothesis
we deduce that $z\in Z_q(p)$.
\bbox

\noindent A further property is that, given $(q,z)\in
D^\omega{\cal N}$, it is possible to find a $p\in P_q(z)$ and to
choose the value of ${\cal H}(q,p)$ simultaneously. This property
will be useful in the following in order to simplify the Hamilton
equations. For that purpose we define, for all $h\in \Bbb{R}$, the
{\bf pseudofiber}:
\[
P^h_q(z):= \{ p\in P_q(z)/ {\cal H}(q,p)=h\}.
\]
We then have:
\begin{lemm}\label{2.2.2.dernierlemme}
For all $(q,z)\in Gr^\omega{\cal N}$ the pseudofiber $P^h_q(z)$ is
a affine subspace\footnote{again in the instance of variational
problems on maps $u:\Bbb{R}^2\longrightarrow \Bbb{R}^2$ and the
multisymplectic manifold $\Lambda^2T^{\star}\Bbb{R}^4$, for any
$(q,z)\in Gr^\omega\R^4$ the pseudofiber $P_q^h(z)\subset
\Lambda^2T^\star\R^4$ is an affine line parallel to $\Bbb{R}\left[
\left( v_1^1v_2^2 -v_1^2v_2^1 \right) dx^1\wedge dx^2 -
\epsilon_{ij}v_\nu^j dy^i\wedge dx^\nu + dy^1\wedge dy^2 \right]$,
where $T(v)=z$. (See also Paragraph 2.2.2.)} of
$\Lambda^nT^*_q{\cal N}$ parallel to $\left( T_zD^n_q{\cal
N}\right)^\perp$. Hence $\hbox{dim }P^h_q(z)=\hbox{dim }P_q(z)-1 =
{(n+k)!\over n!k!}-nk-1$.
\end{lemm}
{\em Proof} --- We first remark that, $\forall q\in {\cal N}$
and $\forall z\in D^\omega_q{\cal N}$, $\omega_q$
belongs to $\left( T_zD^\omega_q{\cal N}\right)^\perp$, because of the definition of $D^\omega_q{\cal N}$.
So $\forall \lambda \in \Bbb{R}$, $\forall p\in P_q(z)$, we deduce from (\ref{2.2.2.p-p}) that
$p+\lambda \omega_q\in P_q(z)$ and thus
\[
\begin{array}{ccl}
{\cal H}(q,p+\lambda \omega_q) & = & \langle z,p+\lambda \omega_q\rangle - L(q,z)\\
& = & {\cal H}(q,p) +\lambda \langle z,\omega_q\rangle = {\cal H}(q,p) +\lambda.
\end{array}
\]
Hence we deduce that $\forall h\in \Bbb{R}$, $\forall p\in P_q(z)$,
$\exists !\lambda\in \Bbb{R}$ such that
\[
{\cal H}(q,p+\lambda\omega_q) = h,
\]
so that $P^h_q(z)$ is non empty. Moreover
if $p_0\in P^h_q(z)$ then $p_1\in P^h_q(z)$ if and only if $p_1-p_0\in
\left( T_zD^\omega_q{\cal N}\right)^\perp \cap z^\perp$, where $z^\perp :=\{p\in \Lambda^nT^*_q{\cal N}/\langle z,p\rangle = 0\}$. In order to conclude
observe that
$\left( T_zD^\omega_q{\cal N}\right)^\perp \cap z^\perp =
\left( T_zD^n_q{\cal N}\right)^\perp$. \bbox

\subsubsection{Critical points}
We now look at critical points of the Lagrangian functional using the above
framework. Instead of the usual approach using jet bundles and contact structure,
we shall derive Hamilton equations directly, without writing the
Euler--Lagrange equation.\\

\noindent First we extend the form $\omega$ on ${\cal M}$ by setting
$\omega\simeq \Pi^*\omega$, where $\Pi:{\cal M}\longrightarrow {\cal N}$ is the
bundle projection, and we define $\widehat{\cal G}^\omega$ to be the set of oriented
$n$-dimensional submanifolds $\Gamma$ of ${\cal M}$, such that $\omega_{|\Gamma}>0$
everywhere. A consequence of this inequality is that the restriction of
the projection $\Pi$ to any $\Gamma\in \widehat{\cal G}^\omega$ is an
embedding into ${\cal N}$: we denote by $\Pi(\Gamma)$ its image. It is clear
that $\Pi(\Gamma)\in {\cal G}^\omega$. Then we can
view $\Gamma$ as (the graph of) a section $q\longmapsto p(q)$
of the pull-back of the bundle ${\cal M}\longrightarrow {\cal N}$ by the
inclusion $\Pi(\Gamma)\subset {\cal N}$.\\

\noindent Second, we define the subclass
$\mathfrak{p}\widehat{\cal G}^\omega\subset \widehat{\cal
G}^\omega$ as the set of $\Gamma\in \widehat{\cal G}^\omega$ such
that, $\forall (q,p)\in \Gamma$, $p\in P_q(T_q\Pi(\Gamma))$ (a
contact condition). [As we will see later it can be viewed as the
subset of $\Gamma\in \widehat{\cal G}^\omega$ which satisfy half
of the Hamilton equations.] And given some $G\in {\cal G}^\omega$,
we denote by $\mathfrak{p}\widehat{G}\subset
\mathfrak{p}\widehat{\cal G}^\omega$ the family of submanifolds
$\Gamma\in \mathfrak{p}\widehat{\cal G}^\omega$ such that
$\Pi(\Gamma)=G$ and we say that $\mathfrak{p}\widehat{G}$ is the
set of {\bf Legendre lifts} of $G$.
We hence have $\mathfrak{p}\widehat{\cal G}^\omega=\cup_{G\in {\cal G}^\omega}\mathfrak{p}\widehat{G}$.\\

\noindent
Lastly, we define the functional on $\widehat{\cal G}^\omega$
\[
{\cal I}[\Gamma]:= \int_\Gamma \theta - {\cal H}\omega.
\]
{\bf Properties of the restriction of ${\cal I}$ to $\mathfrak{p}\widehat{\cal G}^\omega$}
--- First we claim that
\begin{equation}\label{2.2.3.i=l}
{\cal I}[\Gamma] = {\cal L}[G],\quad
\forall G\in {\cal G}^\omega, \forall \Gamma \in \mathfrak{p}\widehat{G}.
\end{equation}
This follows from
\[
\begin{array}{ccl}
\displaystyle \int_\Gamma \theta - {\cal H}\omega & = &
\displaystyle \int_G\langle z_G,p(q)\rangle \omega
- {\cal H}(q,p(q))\omega\\
 & = & \displaystyle \int_G \left( \langle z_G,p(q)\rangle
- \langle z_G,p(q)\rangle + L(q,z_G)\right) \omega = \int_GL(q,z_G)\omega,
\end{array}
\]
where $G\longrightarrow {\cal M}:q\longmapsto (q,p(q))$ is the parametrization of
$\Gamma$ and where $z_G$ is the unique $n$-vector in $D^\omega_q{\cal N}$ (for $q\in G$)
which spans $T_qG$.\\
Second let us exploit relation (\ref{2.2.3.i=l}) to compute the first variation of
${\cal I}$ at any submanifold $\Gamma\in \mathfrak{p}\widehat{G}$, i.e.\,a Legendre lift of
$G\in {\cal G}^\omega$. We let $\xi\in \Gamma({\cal N},T{\cal N})$ be
a smooth vector field with compact support and $G_s$, for $s\in \Bbb{R}$, be
the image of $G$ by the flow diffeomorphism $e^{s\xi}$. For small values of $s$,
$G_s$ is still in ${\cal G}^\omega$ and for all $q_s:=e^{s\xi}(q)\in G_s$
we shall denote by $z_s$ the unique $n$-vector in $D^\omega_{q_s}{\cal N}$ which
spans $T_{q_s}G_s$. Then we choose a smooth section $(s,q_s)\longmapsto p(q)_s$
in such a way that $p(q)_s\in P_{q_s}(z_s)$. This builds a family of Legendre lifts
$\Gamma_s=\{(q_s,p(q)_s)\}$. We can now use relation (\ref{2.2.3.i=l}):
${\cal I}[\Gamma_s]= {\cal L}[G_s]$ and derivate it with respect to $s$.
Denoting by $\widehat{\xi}\in T_{(q,p(q))}{\cal M}$ the vector $d(q_s,p(q)_s)/ds_{|s=0}$, we obtain
\begin{equation}\label{2.2.3.xi}
\delta{\cal I}[\Gamma](\widehat{\xi}) = {d\over ds}{\cal I}[\Gamma_s]_{|s=0} =
{d\over ds}{\cal L}[G_s]_{|s=0} = \delta{\cal L}[G](\xi).
\end{equation}
{\bf Variations of ${\cal I}$ along $T_p{\cal M}_q$} ---
On the other hand for all $\Gamma\in \widehat{\cal G}^\omega$ and
for all {\bf vertical} tangent vector field along $\Gamma$
$\zeta$, i.e.\,such that $d\Pi_{(q,p)}(\zeta) = 0$ or such that
$\zeta\in T_p{\cal M}_q\subset T_{(q,p)}{\cal M}$, we have
\begin{equation}\label{2.2.3.zeta}
\delta{\cal I}[\Gamma](\zeta) = \int_\Gamma \left( \langle z_{\Pi(\Gamma)},\zeta\rangle
- {\partial {\cal H}\over \partial p}(q,p)(\zeta)\right) \omega,
\end{equation}
where $z_{\Pi(\Gamma)}$ is the unique $n$-vector in $D^\omega_q{\cal N}$ (for $q\in G(\Gamma)$)
which spans $T_q\Pi(\Gamma)$. Note that in the special case where
$\Gamma\in \mathfrak{p}\widehat{\cal G}^\omega$,
we have $z_{\Pi(\Gamma)}\in Z_q(p)$, so we deduce from (\ref{2.2.2.hl}) and (\ref{2.2.3.zeta})
that $\delta{\cal I}[\Gamma](\zeta) = 0$. And the converse is true. So $\mathfrak{p}\widehat{\cal G}^\omega$
can be characterized by requiring that condition (\ref{2.2.3.zeta}) is true for all vertical
vector fields $\zeta$.\\

\noindent {\bf Conclusion} ---
The key point is now that any vector field along $\Gamma$ can be written
$\widehat{\xi} + \zeta$, where $\widehat{\xi}$ and $\zeta$ are as above. And
for any $G\in {\cal G}^\omega$ and for all $\Gamma\in \mathfrak{p}\widehat{G}$,
the first variation of ${\cal I}$ at $\Gamma$ with respect to a vector field
$\widehat{\xi} + \zeta$, where locally $\widehat{\xi}$ lifts $\xi\in T_q{\cal N}$
and $\zeta \in  T_p{\cal M}_q$, satisfies
\begin{equation}\label{2.2.3.firstv}
\delta{\cal I}[\Gamma](\widehat{\xi} + \zeta) = \delta{\cal L}[G](\xi).
\end{equation}
We deduce the following.
\begin{theo}\label{2.2.3.equivalence}
(i) For any $G\in {\cal G}^\omega$ and for all Legendre lift $\Gamma\in \mathfrak{p}\widehat{G}$,
$G$ is a critical point of ${\cal L}$ if and only if $\Gamma$ is a critical
point of ${\cal I}$.\\
(ii) Moreover for all $\Gamma\in \widehat{\cal G}^\omega$, if $\Gamma$ is a
critical point of ${\cal I}$ then $\Gamma$ is a Legendre lift,
i.e.\,$\Gamma\in \mathfrak{p}\widehat{\Pi(\Gamma)}$
and $\Pi(\Gamma)$ is a critical point of ${\cal L}$.
\end{theo}
{\em Proof} --- (i) is a straightforward consequence of (\ref{2.2.3.firstv}). Let us
prove (ii): if $\Gamma\in \widehat{\cal G}^\omega$ is a
critical point of ${\cal I}$, then in particular for all vertical tangent vector
field $\zeta\in T_p{\cal M}_q$, $\delta{\cal I}[\Gamma](\zeta)=0$ and
by (\ref{2.2.3.zeta}) this implies $(z_{\Pi(\Gamma)})_{|T_p^*{\cal M}_q} =
(\partial {\cal H}/ \partial p)(q,p)$. Then by applying Lemma \ref{2.2.2.dh}--(ii)
we deduce that $z_{\Pi(\Gamma)}\in Z_q(p)$. Hence $\Gamma$ is a Legendre lift.
Lastly we use the conclusion of the part (i) of the Theorem to conclude
that $G(\Gamma)$ is a critical point of ${\cal L}$. \bbox

\begin{coro}\label{2.2.3.coro1}
Let $\Gamma\in \widehat{\cal G}^\omega$ be a critical point of ${\cal I}$ and let a
smooth section $\pi:\Gamma\longrightarrow \Lambda^nT^*{\cal N}$ satisfy:
$\forall (q,p)\in \Gamma$,
$\pi(q,p)\simeq \pi(q)\in \left( T_zD^\omega_q{\cal N}\right) ^\perp$ (where $z\in Z_q(p)$).
Then $\tilde{\Gamma}:= \{ (q,p+\pi(q))/(q,p)\in \Gamma \}$ is another
critical point of ${\cal I}$.
\end{coro}
{\em Proof} --- By using Theorem \ref{2.2.3.equivalence}--(ii) we deduce that $\Gamma$
has the form $\Gamma=\{(q,p)/q\in \Pi(\Gamma),p\in P_q(z_{\Pi(\Gamma)})\}$ and thus
$\tilde{\Gamma}=\{(q,p+\pi(q))/q\in \Pi(\Gamma),p\in P_q(z_{\Pi(\Gamma)})\}$.  This implies,
by using (\ref{2.2.2.p-p}), that $\tilde{\Gamma}\in \mathfrak{p}\widehat{\Pi(\Gamma)}$;
then $\tilde{\Gamma}$ is also a critical point of ${\cal I}$ because of
Theorem \ref{2.2.3.equivalence}--(i). \bbox

\noindent Note that, for any constant $h\in \Bbb{R}$, by choosing
$\pi(q)=\left( h - {\cal H}(q,p)\right) \omega_q$ (see the proof
of Lemma \ref{2.2.2.dernierlemme}) in the above Corollary we
deform any critical point $\Gamma$ of ${\cal I}$ $\Gamma\in
\widehat{\cal G}^\omega$ into a critical point $\tilde{\Gamma}$ of
${\cal I}$ contained in ${\cal M}^h:=\{m\in {\cal M}/{\cal
H}(m)=h\}$.

\begin{defi}
An {\bf Hamiltonian $n$-curve} is a critical point $\Gamma$ of ${\cal I}$ such that
there exists a constant $h\in \Bbb{R}$ such that $\Gamma\subset {\cal M}^h$...
\end{defi}

\subsubsection{Hamilton equations}
\noindent
We now end this section by looking at the equation satisfied by critical points
of ${\cal I}$. Let $\Gamma\in \widehat{\cal G}^\omega$ and $\xi\in\Gamma({\cal M},
T{\cal M})$ be a smooth vector field with compact support. We let $e^{s\xi}$
be the flow mapping of $\xi$ and $\Gamma_s$ be the image of $\Gamma$ by $e^{s\xi}$.
We let ${\cal X}$ be an $n$-dimensional manifold diffeomorphic to $\Gamma$ and we denote by
\[
\begin{array}{cccl}
\sigma : & (0,1)\times {\cal X} & \longrightarrow & {\cal M}\\
 & (s,x) & \longmapsto & \sigma(s,x)
\end{array}
\]
a map such that if $\gamma_s:x\longmapsto \sigma(s,x)$, then $\gamma=\gamma_0$
is a parametrization of $\Gamma$, $\gamma_s$ is a parametrization of $\Gamma_s$
and ${\partial \over \partial s}\left(\sigma (s,x)\right) = \xi\left(\sigma (s,x)\right)$.
Then
\[
\begin{array}{ccl}
\displaystyle
{\cal I}[\Gamma_s]- {\cal I}[\Gamma] & = &
\displaystyle \int_{\cal X}\gamma_s^*(\theta -{\cal H}\omega)
- \gamma^*(\theta -{\cal H}\omega)\\
 & = & \displaystyle \int_{\partial \left( (0,s)\times {\cal X}\right)}
\sigma^*(\theta -{\cal H}\omega)
= \int_{(0,s)\times {\cal X}}d\left(\sigma^*(\theta -{\cal H}\omega) \right)\\
& = & \displaystyle \int_{(0,s)\times {\cal X}}\sigma^*(\Omega -d{\cal H}\wedge \omega) ).
\end{array}
\]
Thus
\[
\begin{array}{ccl}
\displaystyle
\lim_{s\rightarrow 0}{{\cal I}[\Gamma_s]- {\cal I}[\Gamma]\over s} & = &
\displaystyle \lim_{s\rightarrow 0}{1\over s}\int_{(0,s)\times {\cal X}}
\sigma^*(\Omega -d{\cal H}\wedge \omega)\\
& = & \displaystyle \int_{\cal X}{\partial \over \partial s}\iN
\sigma^*(\Omega -d{\cal H}\wedge \omega)
= \int_{\cal X}\gamma^*(\xi\iN (\Omega -d{\cal H}\wedge \omega) )\\
& = & \displaystyle \int_\Gamma \xi\iN (\Omega -d{\cal H}\wedge \omega) .
\end{array}
\]
We hence conclude that $\Gamma$ is a critical point of ${\cal I}$ if and only if
$\forall m\in \Gamma$, $\forall \xi\in T_m{\cal M}$, $\forall X\in \Lambda^nT_m\Gamma$,
\[
\xi\iN (\Omega -d{\cal H}\wedge \omega)(X) = 0
\quad \Longleftrightarrow \quad
X\iN (\Omega -d{\cal H}\wedge \omega)(\xi) = 0.
\]
We thus deduce the following.
\begin{theo}\label{2.2.4.theo}
A submanifold $\Gamma\in \widehat{\cal G}^\omega$ is a critical point of ${\cal I}$
if and only if
\begin{equation}\label{2.2.4.hamilton-}
\forall m\in \Gamma, \forall X\in \Lambda^nT_m\Gamma,
\quad X\iN (\Omega -d{\cal H}\wedge \omega) = 0.
\end{equation}
Moreover, if there exists some $h\in \Bbb{R}$ such that $\Gamma\subset {\cal M}^h$
(i.e.\,$\Gamma$ is a Hamiltonian $n$-curve) then
\begin{equation}\label{2.2.4.hamilton}
\forall m\in \Gamma, \exists ! X\in \Lambda^nT_m\Gamma,
\quad X\iN \Omega = (-1)^nd{\cal H}.
\end{equation}
\end{theo}
Recall that,
because of Lemma \ref{2.2.2.dernierlemme} and Corollary \ref{2.2.3.coro1}, it is always
possible to deform a Hamiltonian $n$-curve $\Gamma\longmapsto \tilde{\Gamma}$
in such a way that ${\cal H}$ be constant on $\tilde{\Gamma}$ and
$\Pi(\Gamma) = \Pi(\tilde{\Gamma})$.\\
{\em Proof} --- We just need to check (\ref{2.2.4.hamilton}). Let $\Gamma\subset {\cal M}^h$.
Since $d{\cal H}_{|\Gamma}=0$, $\forall X\in \Lambda^nT_m\Gamma$,
$X\iN d{\cal H}\wedge \omega = (-1)^n\langle X,\omega\rangle d{\cal H}$. So by choosing
the unique $X$ such that $\langle X,\omega\rangle =1$, we obtain
$X\iN d{\cal H}\wedge \omega = (-1)^nd{\cal H}$. Then (\ref{2.2.4.hamilton-}) is
equivalent to (\ref{2.2.4.hamilton}).\bbox

\subsection{Some examples}
We pause to study on some simple examples how the Legendre correspondence and the
Hamilton work. In particular in the construction of ${\cal M}$ we let a large freedom
in the dimension of the fibers ${\cal M}_q$, having just the constraint
that $\dim {\cal M}_q\leq \dim {\cal P}_q={(n+k)!\over n!k!}$. This leads to a large choice
of approaches between two opposite ones: the first one consists in using as less variables as
possible, i.e.\,to choose ${\cal M}$ to be of minimal dimension
(for example the de Donder--Weyl theory),
the other one consists in using the largest number of
variables, i.e.\,to choose ${\cal M}$ to be equal to the interior of ${\cal P}$ (the advantage
will be that in some circumstances we avoid degenerate situations).\\

\noindent We focus here on special cases of Example 2 of the previous Section:
we consider maps $u:{\cal X}\longrightarrow {\cal Y}$. We denote by $q^\mu=x^\mu$,
if $1\leq \mu\leq n$, coordinates on ${\cal X}$ and by $q^{n+i}=y^i$, if $1\leq i\leq k$,
coordinates on ${\cal Y}$. Recall that $\forall x\in {\cal X}$, $\forall y\in {\cal Y}$,
the set of linear maps $v$ from $T_x^*{\cal X}$ to $T_y{\cal Y}$ can be identified with
$T_y{\cal Y}\otimes T_x^*{\cal X}$. And coordinates representing some
$v\in T_y{\cal Y}\otimes T_x^*{\cal X}$ are denoted by $v^i_\mu$, in such a way that
$v=\sum_\alpha\sum_i v^i_\mu{\partial \over \partial y^i}\otimes dx^\mu$.
Then through the diffeomorphism
$T_y{\cal Y}\otimes T_x^*{\cal X}\ni v\longmapsto T(v)\in  Gr^\omega_{(x,y)}{\cal N}$
(where ${\cal N}={\cal X}\times {\cal Y}$)
we obtain coordinates on $Gr^\omega_q{\cal N}\simeq D^\omega_q{\cal N}$.
We also denote by $e:=p_{1\cdots n}$, $p^{\mu}_i:=p_{1\cdots (\mu-1)i(\mu+1)\cdots n}$,
$p^{\mu_1\mu_2}_{i_1i_2}:=p_{1\cdots (\mu_1-1)i_1(\mu_1+1)\cdots
(\mu_2-1)i_2(\mu_2+1)\cdots n}$, etc., so that
\[
\Omega=de\wedge \omega + \sum_{j=1}^n\sum_{\mu_1<\cdots <\mu_j}
\sum_{i_1<\cdots <i_j}dp^{\mu_1\cdots \mu_j}_{i_1\cdots i_j}\wedge
\omega_{\mu_1\cdots \mu_j}^{i_1\cdots i_j},
\]
where, for $1\leq p\leq n$,
\[
\begin{array}{ccl}
\omega & := & dx^1\wedge \cdots  \wedge dx^n\\
\omega^{i_1\cdots i_p}_{\mu_1\cdots \mu_p} & := & dy^{i_1}\wedge \cdots  \wedge
dy^{i_p}\wedge \left( {\partial \over \partial x^{\mu_1}}\wedge \cdots  \wedge
{\partial \over \partial x^{\mu_p}}\iN \omega\right) .
\end{array}
\]
{\bf Remark} --- It can be checked (see for instance \cite{HeleinKouneiher}) that,
by denoting by $p^*$ all coordinates
$p^{\mu_1\cdots \mu_j}_{i_1\cdots i_j}$ for $j\geq 1$, the Hamiltonian function has
always the form ${\cal H}(q,e,p^*) = e+H(q,p^*)$.

\subsubsection{The de Donder--Weyl formalism}
In the special case of the de Donder--Weyl theory, ${\cal M}^{dDW}_q$ is the submanifold of
$\Lambda^nT^*_q{\cal N}$ defined by the constraints
$p^{\mu_1\cdots \mu_j}_{i_1\cdots i_j} = 0$, for all $j\geq 2$
(Observe that these constraints are invariant by a change of coordinates, so that
they have an intrinsic meaning.) We thus have
\[
\Omega^{dDW}=de\wedge \omega + \sum_\mu \sum_idp^\mu_i\wedge \omega_\mu^i..
\]
Then the equation $\partial W/ \partial z(q,z,p) = 0$ is
equivalent to $p^\mu_i=\partial l/ \partial v^i_\mu(q,v)$, so that
the Legendre Correspondence Hypothesis holds if and only if
$(q,v)\longmapsto (q,\partial l /\partial v(q,p))$ is an
invertible map. Note that then the enlarged pseudofibers $P_q(z)$
intersect ${\cal M}^{dDW}_q$ along lines $\{e\omega + {\partial
l\over \partial v^i_\mu}(q,v)\omega_\mu^i/e\in \Bbb{R}\}$. So
since dim$\Lambda^nT^*_q{\cal N}= {(n+k)!\over n!k!}$, dim${\cal
M}^{dDW}_q=nk+1$ and dim$P_q(z)={(n+k)!\over n!k!}-nk$, the
Legendre Correspondence Hypothesis can be rephrased by saying that
each $P_q(z)$ meets ${\cal M}^{dDW}_q$ transversally along a line.
Moreover $Z_q(e\omega + p^\mu_i\omega_\mu^i)$ is then reduced to
one point, namely $T(v)$, where
$v$ is the solution to $p^\mu_i={\partial l\over \partial v^i_\mu}(q,v)$.\\

\noindent For more details and a description using local coordinates, see
\cite{HeleinKouneiher}.

\subsubsection{Maps from $\Bbb{R}^2$ to $\Bbb{R}^2$ via the Lepage--Dedecker point of view}
Let us consider a simple situation where ${\cal X} = {\cal Y} = \Bbb{R}^2$ and
${\cal M}\subset \Lambda^2T^{\star}\Bbb{R}^4$. It corresponds to variational problems on maps
$u:\Bbb{R}^2\longrightarrow \Bbb{R}^2$. For any point $(x,y)\in \Bbb{R}^4$, we denote
by $(e,p^i_\mu,r)$ the coordinates on $\Lambda^2T_{(x,y)}\Bbb{R}^4$, such that
$\theta = e\,dx^1\wedge dx^2 + p^1_idy^i\wedge dx^2 + p^2_idx^1\wedge dy^i + r\,dy^1\wedge dy^2$.
An explicit parametrization of $\{z\in D^2_{(x,y)}\Bbb{R}^4/\omega(z)>0\}$ is given by the coordinates
$(t,v^i_\mu)$ through
\[
z = t^2{\partial \over \partial x^1}\wedge {\partial \over \partial x^2} +
t\,\epsilon^{\mu\nu} v^i_\mu{\partial \over \partial y^i}\wedge {\partial \over \partial x^\nu}
+ (v^1_1v^2_2-v^1_2v^2_1){\partial \over \partial y^1}\wedge {\partial \over \partial y^2},
\]
where $\epsilon^{12} = - \epsilon^{21} =1$ and $\epsilon^{11} = \epsilon^{22} = 0$.
One then finds that $\left(T_zD^2_q\Bbb{R}^4\right)^\perp$ is\\
$\Bbb{R}\left[
\left( v_1^1v_2^2 -v_1^2v_2^1 \right) dx^1\wedge dx^2 -
\epsilon_{ij}v_\nu^j dy^i\wedge dx^\nu + dy^1\wedge dy^2
\right]$, whereas $\left(T_zD^\omega_q\Bbb{R}^4\right)^\perp$ is
$\left(T_zD^2_q\Bbb{R}^4\right)^\perp \oplus \Bbb{R}dx^1\wedge dx^2$.\\

\noindent
We deduce that the sets $P_q(z)$ and $P^h_q(z)$ form a family
of non parallel affine subspaces so we expect that on the one hand these subspaces will
intersect, causing obstructions there for the invertibility of the Legendre mapping,
and on the other hand they will fill ``almost'' all of $\Lambda^2T^{\star}_{(x,y)}\Bbb{R}^4$,
giving rise to the phenomenon that the Legendre correspondence is ``generically
everywhere'' well defined.\\

\noindent {\bf Example 4} --- {\em The trivial variational problem} ---
We just take $l =0$, so that any map map from $\Bbb{R}^2$ to $\Bbb{R}^2$
is a critical point of $\ell$ ! This example is motivated by gauge theories where the
gauge invariance gives rise to constraints. In this case the sets $P_q(z)$ are
exactly $\left(T_zD^\omega_q\Bbb{R}^4\right)^\perp$ and $\cup_zP_q(z)$ is equal to
${\cal P}_q:=\{(e,p^\mu_i,r)\in \Lambda^2T^*_q\Bbb{R}^4/
r\neq 0 \}\cup \{(e,0,0)/e\in \Bbb{R}\}$. If we assume that $r\neq 0$ and
choose ${\cal M}_q= \{(e,p^\mu_i,r)\in \Lambda^2T^*_q\Bbb{R}^4/r\neq 0\}$,
then
\[
{\cal H}(q,p) = e - {p^1_1p^2_2 - p^1_2p^2_1\over r}.
\]
One can then check that all Hamiltonian 2-curves are of the form
\[
\Gamma=\left\{\left(x,u(x),e(x)dx^1\wedge dx^2+ \epsilon_{\mu\nu}p^\mu_i(x)dy^i\wedge dx^\nu +
r(x) dy^1\wedge dy^2 \right)/x\in \Bbb{R}^2 \right\},
\]
where $u:\Bbb{R}^2\longrightarrow \Bbb{R}^2$ is an {\em arbitrary} smooth function,
$r:\Bbb{R}^2\longrightarrow \Bbb{R}^*$ is also an arbitrary smooth function,
$e(x) = r(x)\left( {\partial u^1\over \partial x^1}(x){\partial u^2\over \partial x^2}(x)
- {\partial u^1\over \partial x^2}(x){\partial u^1\over \partial x^2}(x)\right) + h$,
(for some constant $h\in \Bbb{R}$) and $p^\mu_i(x) = -r(x)\epsilon_{ij}\epsilon^{\mu\nu}
{\partial u^j\over \partial x^\nu}(x)$.\\

\noindent {\bf Example 5} --- {\em The elliptic Dirichlet integral
(see also \cite{HeleinKouneiher})} --- The Lagrangian is $l
(x,y,v) = {1\over 2}|v|^2 + B(v_1^1v_2^2 -v_1^2v_2^1)$ where\footnote{There $B$ could be interpreted as a $B$-field of a bosonic string theory.} $|v|^2:=(v^1_1)^2 + (v^1_2)^2
+(v^2_1)^2 +(v^2_2)^2$. We then find that
\[
{\cal H}(q,p) = e + {1\over 1-(r-B)^2}
\left( {|p|^2\over 2} + (r-B)(p^1_1p^2_2 - p^1_2p^2_1)\right) .
\]
\noindent {\bf Example 6} --- {\em Maxwell equations in two dimensions} ---
We choose $l(x,y,v) = -{1\over 2}\left( v^1_2-v^2_1\right)^2$, so that by
identifying $(u^1,u^2)$ with the components $(A_1,A_2)$ of a
Maxwell gauge potential, we recover the usual Lagrangian
$l (dA)=-{1\over 4}\sum_{\mu,\nu}\left(
{\partial A_\nu\over \partial x^\mu} - {\partial A_\mu\over \partial x^\nu}\right)^2$
for Maxwell fields without charges. We then obtain
\[
{\cal H}(q,p) = e + {(p^1_2 + p^2_1)^2 - 4p^1_1p^2_2\over 4r}
-{1\over 4} {(p^1_2 - p^2_1)^2\over 2+r}.
\]

\noindent {\bf Conclusion} --- It is worth to look at the
differences between the Lepage--Dedecker and the de Donder--Weyl
theories through these examples. Indeed the de Donder--Weyl theory
can be simply recovered by letting $r=0$. One sees immediately
that for the trivial variational problem this forces $p^1_1p^2_2 -
p^1_2p^2_1$ to be 0: actually a more careful inspection shows that
all pseudofibers intersect along $p^\mu_i = 0$ so that all these
components must be set to 0 in the de Donder--Weyl theory.  In the
example of the elliptic Dirichlet functional no constraint
appears unless $B=\pm 1$. And for the Maxwell equations all pseudofibers intersect
along the subspace $p^1_2+p^2_1 = p^1_1 = p^2_2 = 0$ and so we
recover the constraints  already
observed in \cite{Kanatchikov1} and \cite{HeleinKouneiher}
in the de Donder--Weyl formulation.

\subsection{Invariance properties along pseudofibers}
We have seen that for all $q\in {\cal N}$, for $h\in \Bbb{R}$ and $z\in D^\omega_q{\cal N}$,
the pseudofiber $P_q^h(z)$ is an affine subspace of $\Lambda^nT^*_q{\cal N}$ parallel to
$\left(T_zD^n_q{\cal N}\right)^\perp$. Let us assume that ${\cal M}_q$ is an open subset
of $\Lambda^nT^*_q{\cal N}$: then the Legendre Correspondence
Hypothesis implies that $\forall (p,q)\in {\cal M}$, $Z_q(p)$ is reduced to one point that we shall denote by
$Z(q,p)$. Hence we can define the distribution of subspaces on ${\cal M}$ by:
\[
\forall (q,p)\in {\cal M}, \quad L^{\cal H}_{(q,p)}:= \left(T_{Z(q,p)}D^n_q{\cal N}\right)^\perp.
\]
It is actually the subspace tangent to the pseudo-fiber passing through $(q,p)$.
In Section 3.3 we will propose a generalization of the definition of $L^{\cal H}_{(q,p)}$
which makes sense on an arbitrary multisymplectic manifold. We will prove in Section 4.3
that this generalized definition coincides with the first one in the case where the
multisymplectic manifold is $\Lambda^nT^*{\cal N}$.
Lastly Lemma \ref{2.2.2.dernierlemme} and Corollary \ref{2.2.3.coro1} can be rephrased as

\begin{theo}\label{2.3.theoreform}
Let  ${\cal M}$ be an open subset of $\Lambda^nT^{\star}{\cal N}$ and let ${\cal H}$ be a Legendre image  Hamiltonian function on ${\cal M}$  (by means of the
Legendre correspondence). Then
\begin{equation}\label{2.3.dh=0}
\forall (q,p)\in {\cal M},
\forall \xi\in L^{\cal H}_{(q,p)},\quad  d{\cal H}_{(q,p)}(\xi) = 0.
\end{equation}
And if $\Gamma\in \widehat{\cal G}^\omega$ is a Hamiltonian $n$-curve and if
$\xi$ a vector field which is a smooth section of $L^{\cal H}$, then denoting by
$e^{s\xi}$ the flow mapping of $\xi$
\begin{equation}\label{2.3.stability}
\forall s\in \Bbb{R},\hbox{ small enough },
e^{s\xi}(\Gamma)\hbox{ is a Hamiltonian }n{-curve}.
\end{equation}
\end{theo}

\subsection{Gauge theories}
The above theory can be adapted for variational theories on gauge
fields (connections) by using a local trivialization. More precisely, given
a $\mathfrak{g}$-connection $\nabla^0$ acting on a trivial bundle
with structure group $\mathfrak{G}$ (and Lie algebra
$\mathfrak{g}$) any other connection $\nabla$ can be identified
with the $\mathfrak{g}$-valued 1-form $A$ on the base manifold
${\cal X}$ such that $\nabla = \nabla^0 + A$. We may couple $A$ to
a Higgs field $\varphi:{\cal X}\longrightarrow \Phi$, where $\Phi$
is a vector space on which $\mathfrak{G}$ is acting. Then any
choice of a field $(A,\varphi)$ is equivalent to the data of an
$n$-dimensional submanifold $\Gamma$ in ${\cal M}:=
\left(\mathfrak{g}\otimes T^*{\cal X}\right)\times \Phi$ which is
a section of this fiber bundle over ${\cal X}$. An example of this
approach
is the one that we use for the Maxwell field at the end of this paper.\\

\noindent
But if we wish to study more general gauge theories and in particular
connections on a non trivial bundle we need a more general and more covariant framework.
Such a setting can consist in viewing a connection as a $\mathfrak{g}$-valued
1-form $a$ on a principal bundle ${\cal F}$ over the space-time satisfying some
equivariance conditions (under some action of the group $\mathfrak{G}$).
Similarly the Higgs field, a section of an associated bundle, can
be viewed as an equivariant map $\phi$ on ${\cal F}$ with values in a fixed space.
Thus the pair $(a,\phi)$ can be pictured geometrically as a section $\Gamma$, i.e.\,a submanifold
of some fiber bundle ${\cal N}$ over ${\cal F}$, satisfying two kinds of constraints:
\begin{itemize}
\item
$\Gamma$ is contained in a submanifold ${\cal N}_{\mathfrak{g}}$ (a geometrical
translation of the constraints ``the restriction of $a_f$ to the subspace tangent to the
fiber ${\cal F}_f$ is $-dg\cdot g^{-1}$'') and
\item $\Gamma$ is invariant
by an action of $\mathfrak{G}$ on ${\cal N}$ which preserves ${\cal N}_{\mathfrak{g}}$.
\end{itemize}
Within this more abstract framework we are reduced to a situation similar to the one
studied in the beginning of this Section, but we need to understand what are the consequence
of the two equivariance conditions. (In particular this will imply that there is a canonical
distribution of subspaces which is tangent to all pseudofibers).
This will be done in details in \cite{HeleinKouneiher1.1}.
In particular we compare this abstract point of view with the more naive one expounded
above.

\section{Multisymplectic manifolds}

We now set up a general framework extending the situation encountered in the previous Section.
\subsection{Definitions}
Recall that, given a differential manifold ${\cal M}$ and $n\in
\Bbb{N}$ a smooth $(n+1)$-form $\Omega$ on ${\cal M}$ is a {\bf
multisymplectic} form if and only if (i) $\Omega$ is non
degenerate, i.e.\,$\forall m\in {\cal M}$, $\forall \xi \in
T_m{\cal M}$, if $\xi \iN \Omega_m = 0$, then $\xi = 0$ (ii)
$\Omega$ is closed, i.e.\,$d\Omega = 0$. And we call any manifold
${\cal M}$ equipped with a multisymplectic form $\Omega$ a {\bf
multisymplectic} manifold. (See Definition \ref{0.def1}.) In the
following, $N$ denotes the dimension of ${\cal M}$. For any $m\in
{\cal M}$ we define the set
\[
D^n_m{\cal M}:= \{X_1\wedge \cdots \wedge X_n\in \Lambda^nT_m{\cal M}/
X_1,\cdots , X_n\in T_m{\cal M}\},
\]
of {\bf decomposable} $n$-vectors and denote by $D^n{\cal M}$ the
associated bundle.

\begin{defi}\label{2.1.def2}
Let ${\cal H}$ be a smooth real valued function defined over a
multisymplectic manifold $({\cal M}, \Omega)$. A Hamiltonian
$n$-curve $\Gamma$ is a $n$-dimensional submanifold of ${\cal M}$
such that for any $m\in \Gamma$, there exists a $n$-vector $X$ in
$\Lambda^nT_m\Gamma$ which satisfies
$$X\iN \Omega = (-1)^nd{\cal H}.$$
We denote by ${\cal E}^{\cal H}$ the set of all such Hamiltonian $n$-curves. We shall
also write for all $m\in {\cal M}$, $[X]^{\cal H}_m:=\{X\in D^n_m{\cal M}/X\iN \Omega = (-1)^nd{\cal H}_m\}$.
\end{defi}
A Hamiltonian $n$-curve is automatically oriented by the
$n$-vector $X$ involved in the Hamilton equation. Remark also that
it may happen that no Hamiltonian $n$-curve exist. An example is
${\cal M}:=\Lambda^2T^{\star}\Bbb{R}^4$ with $\Omega =
\sum_{1\leq\mu<\nu\leq 4}dp_{\mu\nu}\wedge dq^\mu\wedge dq^\nu$
for the case ${\cal H}(q,p)=p_{12}+p_{34}$. Assume that a
Hamiltonian 2-curve $\Gamma$ would exist and let
$X:(t^1,t^2)\longmapsto X(t^1,t^2)$ be a parametrization of
$\Gamma$ such that ${\partial X\over \partial t^1}\wedge {\partial
X\over \partial t^2}\iN \Omega = (-1)^2d{\cal H}$. Then, denoting
by $X_\mu:={\partial X\over \partial t^\mu}$, we would have
$dx^\mu\wedge dx^\nu(X_1,X_2) = {\partial {\cal H}\over \partial
p_{\mu\nu}}$, which is equal to $\pm 1$ if $\{\mu,\nu\}=\{1,2\}$
or $\{3,4\}$ and to 0 otherwise. But this would contradict the
fact that $X_1\wedge X_2$ is decomposable. Hence there is no
Hamiltonian
2-curve in this case.\\

\noindent
Note that beside the the Lepage--Dedecker multisymplectic
manifold $(\Lambda^nT^*{\cal N}, \Omega)$ studied in the previous Section, other examples
of multisymplectic manifolds arises naturally as for example a multisymplectic structure
associated to the Palatini formulation of pure gravity in
4-dimensional space-time (see \cite{HK1}, \cite{HK1b}, \cite{Rovelli}).\\

\noindent In the following we address questions related to the
following general problematic, set in the spirit of the general
relativity: assume that a field theory (and in particular
including a space-time description) is modelled by a
multisymplectic manifold $({\cal M}, \Omega)$ and possibly a
Hamiltonian ${\cal H}$. How could we recover its physical
properties, i.e.\,understand how space-time coordinates merge out,
how momenta and energy appear, without using {\em ad hoc}
hypotheses ? We probably do not know enough to be able to answer
such questions and in the following we will content ourself with
partial answers.

\subsection{The notion of $r$-regular functions}
This question is motivated by the search for understanding space-time coordinates. One could characterize
components of a space-time chart as functions which: (i) are defined for all possible dynamics, (ii)
allow us to separate any pair of different points on space-time. The easiest way to fulfill the
first requirement is to assume that any coordinate function is obtained as the restriction of
a function $f:{\cal M}\longrightarrow \R$ on the Hamiltonian $n$-curve describing the dynamics.
The infinitesimal version of the second requirement is then to assume that the restriction of
the $n$ functions chosen $f^1,\cdots ,f^n$ on any Hamiltonian $n$-curve is locally a diffeomorphism.
This motivates the following

\begin{defi}\label{3.2.defi-reg}
Let $({\cal M},\Omega)$ be a multisymplectic manifold and ${\cal
H}\in {\cal C}^\infty({\cal M})$ a Hamiltonian function. Let
$1\leq r\leq n$ be an integer. A function $f\in {\cal C}^1({\cal
M},\R^r)$ is called {\bf $r$-regular} if and only if for any
Hamiltonian $n$-curve $\Gamma\subset {\cal M}$ the restriction
$f_{|\Gamma}$ is a submersion.
\end{defi}
The dual notion is:
\begin{defi}\label{2.1.def40}
Let ${\cal H}$ be a smooth real valued function defined over a
multisymplectic manifold $({\cal M}, \Omega)$. A {\bf slice of
codimension} $r$ is a cooriented submanifold $\Sigma$ of ${\cal
M}$ of codimension $r$ such that for any $\Gamma\in {\cal E}^{\cal
H}$, $\Sigma$ is transverse to $\Gamma$. By {\bf cooriented} we
mean that for each $m\in \Sigma$, the quotient space $T_m{\cal
M}/T_m\Sigma$ is oriented continuously in function of $m$.
\end{defi}
Indeed it is clear that the level sets of a $r$-regular function $f:{\cal M}\longrightarrow \R^r$
are slices of codimension $r$.\\
{\bf Example 7} --- {\em The case when ${\cal M} =
\Lambda^nT^{\star}({\cal X}\times {\cal Y})$ and that ${\cal
H}(x,y,p) = e + H(x,y,p^*)$ as in Section 2.2} ---Let $\Pi_{\cal
X}:{\cal M}\longrightarrow {\cal X}$ be the natural projection.
Then for any function $\varphi\in {\cal C}^1({\cal X},\R^r)$
without critical point (i.e.\,$d\varphi$ is of rank $r$
everywhere) the function $\varphi\circ \Pi_{\cal X}:{\cal
M}\longrightarrow \R^r$ a $r$-regular function. Indeed because of
the particular dependance of ${\cal H}$ on $e$ a Hamiltonian
$n$-curve is always a graph over ${\cal X}$. A particular case is
when $r=1$, then  any level set $\Sigma$ of $\varphi$ is a
codimension 1 slice and a (class of) vector $\tau\in T_m{\cal
M}/T_m\Sigma$
is positively oriented if and only if $d\varphi(\tau)>0$.\\

\noindent
Note that in this framework an event in space-time can be represented by a slice of codimension $n$. The notion of slice is also important because it helps to construct observable
functionals on the set of solutions ${\cal E}^{\cal H}$. Indeed if $F$ is a $(n-1)$-form
on ${\cal M}$ and if $\Sigma$ is a slice of codimension $1$ we define the
{\em functional} denoted symbolically by
$\int_{\Sigma}F:{\cal E}^{\cal H}\longmapsto \Bbb{R}$ by:
\[
\Gamma\quad \longmapsto \int_{\Sigma\cap \Gamma} F.
\]
Here the intersection $\Sigma\cap \Gamma$ is oriented as follows:
assume that $\alpha\in T_m^{\star}{\cal M}$ is such that
$\alpha_{|T_m\Sigma}=0$ and $\alpha>0$ on $T_m{\cal M}/T_m\Sigma$
and let $X\in \Lambda^nT_m\Gamma$ be positively oriented. Then we
require that $X\Ni \alpha\in \Lambda^{n-1}T_m(\Sigma\cap \Gamma)$
is positively oriented. We can further assume restrictions on the
choice of $F$ in order to guarantee the fact that the resulting
functional is physically observable. Such a situation is achieved
if for example $F$ is so that $dF_{|T_m\Gamma}$ depends only
on $d{\cal H}_m$ (see \cite{HK1b} for details).\\

\noindent In the next Section we will study a characterization of
$r$-regular functions in the special case where ${\cal M} =
\Lambda^nT^*{\cal N}$.

\subsection{Pataplectic invariant Hamiltonian functions}
In Section 2.3 we gave a definition of the subspaces tangent
to the pseudofibers $L^{\cal H}_m$ which was directly deduced from our
analysis of pseudofibers. In Section 4.3 we will prove that an
alternative characterization of $L^{\cal H}_m$ in
$\Lambda^nT^*{\cal N}$ exists and is more intrinsic. It motivates
the following definition: given an arbitrary multisymplectic
manifold $({\cal M}, \Omega)$ and a Hamiltonian function ${\cal
H}:{\cal M}\longrightarrow \Bbb{R}$ and for all $m\in {\cal M}$ we
define the {\bf generalized pseudofiber direction} to be
\begin{equation}\label{3.3.4.lh}
\begin{array}{ccl}
L^{\cal H}_m & := & \displaystyle \left(T_{[X]_m^{\cal H}}D^n_m{\cal M}\iN \Omega\right)^{\perp}\\
 & := & \displaystyle \{\xi \in T_m{\cal M}/\forall X\in [X]^{\cal H}_m,
\forall \delta X\in T_XD^n_m{\cal M}, \xi\iN \Omega(\delta X) = 0\}.
\end{array}
\end{equation}
And we write $L^{\cal H}:=\cup_{m\in {\cal M}}L^{\cal H}_m\subset T{\cal M}$ for
the associated distribution of subspaces.\\

\noindent
Note that if we choose an arbitrary Hamiltonian function ${\cal H}$,
there is no reason for the conclusions of Theorem \ref{2.3.theoreform} to be true, unless we
know that ${\cal H}$ was created out of a Legendre correspondence. This motivates the
following definition\footnote{ In the following if $\xi$ is a
smooth vector field, we denote by $e^{s\xi}$ (for $s\in I$, where $I$ is an
interval of $\Bbb{R}$) its
flow mapping. And if $E$ is any subset of ${\cal M}$, we denote by
$E_s:=e^{s\xi}(E)$ its image by $e^{s\xi}$.}:
\begin{defi}\label{3.3.4.def}
We say that ${\cal H}$ is {\bf pataplectic invariant} if
\begin{enumerate}
\item $\forall \xi\in L^{\cal H}_m$, $d{\cal H}_m(\xi)=0$
\item for all Hamiltonian $n$-curve $\Gamma\in {\cal E}^{\cal H}$, for all
vector field $\xi$ which is a smooth section of $L^{\cal H}$, then,
for $s\in \Bbb{R}$ sufficiently small, $\Gamma_s:=e^{s\xi}(\Gamma)$ is also
a Hamiltonian $n$-curve.
\end{enumerate}
\end{defi}
In \cite{HK1b} we prove that, if ${\cal H}$ is pataplectic
invariant and if some further hypotheses are fulfilled,
functionals of the type $\int_\Sigma F$ are invariant by
deformations along $L^{\cal H}$.

\section{The study of $\Lambda^nT^{\star}{\cal N}$}
In this Section we analyze in details the special case where
${\cal M}$ is an open subset of $\Lambda^nT^{\star}{\cal N}$. Since we are
interested here in local properties of ${\cal M}$, we will use local
coordinates $m=(q,p)=(q^\alpha,p_{\alpha_1\cdots \alpha_n})$ on ${\cal M}$, and the
multisymplectic form reads $\Omega=\sum_{\alpha_1<\cdots <\alpha_n}dp_{\alpha_1\cdots \alpha_n}\wedge
dq^{\alpha_1}\wedge \cdots  \wedge dq^{\alpha_n}$.
For $m=(q,p)$, we write
\[
d_q{\cal H}:= \sum_{1\leq \alpha\leq n+k}{\partial {\cal H}\over \partial q^\alpha}dq^\alpha,\quad
d_p{\cal H}:= \sum_{1\leq \alpha_1<\cdots <\alpha_n\leq n+k}
{\partial {\cal H}\over \partial p_{\alpha_1\cdots \alpha_n}}dp_{\alpha_1\cdots \alpha_n},
\]
so that $d{\cal H} = d_q{\cal H} + d_p{\cal H}$.
\subsection{The structure of $[X]^{\cal H}_m$}
Here we are given some Hamiltonian function ${\cal H}:{\cal
M}\longrightarrow \Bbb{R}$ and a point $ m\in {\cal M}$ such that
$[X]^{\cal H}_m\neq \emptyset$ and\footnote{observe that, although
the splitting $d{\cal H} = d_q{\cal H} + d_p{\cal H}$ depends on a
trivialization of $\Lambda^nT^*{\cal N}$, the condition $d_p{\cal
H}_m \neq 0$ is intrinsic: indeed it is equivalent to $d{\cal
H}_{m|\hbox{Ker}d\Pi_m} \neq 0$, where $\Pi:\Lambda^nT^*{\cal N}
\longrightarrow {\cal N}$.} $d_p{\cal H}_m\neq 0$. Given any
$X=X_1\wedge \cdots \wedge X_n\in D^n_m{\cal M}$ and any form
$a\in T^*_m{\cal M}$ we will write that $a_{|X}\neq 0$ (resp.
$a_{|X}= 0$) if and only if $(a(X_1),\cdots ,a(X_n))\neq 0$ (resp.
$(a(X_1),\cdots ,a(X_n))= 0$). We will say that a form $a\in
T^*_m{\cal M}$ is {\bf proper on} $[X]^{\cal H}_m$ if and only if
it's  either a {\em point-slice}
\begin{equation}\label{4.1/2.differe}
\forall X\in [X]^{\cal H}_m,\quad a_{|X} \neq 0,
\end{equation}
or a {\em co-isotropic}
\begin{equation}\label{4.1/2.nul}
\forall X\in [X]^{\cal H}_m,\quad a_{|X} = 0.
\end{equation}
 We are interested in characterizing all proper
1-forms on $[X]^{\cal H}_m$. We show in this section the
following.
\begin{lemm}\label{4.1/2.lemme}
Let ${\cal M}$ be an open subset of $\Lambda^nT^{\star}{\cal N}$ endowed with its
standard multisymplectic form $\Omega$, let ${\cal H}:{\cal M}\longrightarrow \Bbb{R}$
be a smooth Hamiltonian function. Let $m\in {\cal M}$ such that $d_p{\cal H}_m\neq 0$ and
$[X]^{\cal H}_m\neq \emptyset$. Then\\
(i) the $n+k$ forms $dq^1,\cdots ,dq^{n+k}$
are proper on $[X]^{\cal H}_m$ and satisfy the following property:
$\forall X\in [X]^{\cal H}_m$ and
for all $Y,Z\in T_m{\cal M}$ which are in the vector space spanned by $X$,
if $dq^\alpha(Y) = dq^\alpha(Z)$, $\forall \alpha=1,\cdots ,n+k$, then $Y=Z$.\\
(ii) Moreover for all $a\in T^*_m{\cal M}$ which is proper on $[X]^{\cal H}_m$ we have
\begin{equation}\label{4.1/2.base}
\exists ! \lambda\in \Bbb{R},
\exists !(a_1,\cdots ,a_{n+k})\in \Bbb{R}^{n+k},\quad
a=\lambda d{\cal H}_m + \sum_{\alpha=1}^{n+k}a_\alpha dq^\alpha.
\end{equation}
(iii) Up to a change of coordinates on ${\cal N}$ we can assume that $dq^1,\cdots ,dq^n$
are point-slices and that $dq^{n+1},\cdots ,dq^{n+k}$ satisfy (\ref{4.1/2.nul}). Then
$a\in T^*{\cal M}$ is a point-slice if and only if (\ref{4.1/2.base}) occurs
with $(a_1,\cdots ,a_n)\neq 0$.

\end{lemm}
{\em Proof} --- {\em First step --- analysis of $[X]^{\cal H}_m$}.
We start by introducing some extra notations:
each vector $Y\in T_m{\cal M}$ can be decomposed
into a ``vertical'' part $Y^V$ and a ``horizontal'' part $Y^H$ as follows: for any
$Y=\sum_{1\leq \alpha\leq n+k}Y^\alpha {\partial \over \partial q^\alpha} +
\sum_{1\leq \alpha_1<\cdots <\alpha_n\leq n+k}
Y_{\alpha_1\cdots \alpha_n}{\partial \over \partial p_{\alpha_1\cdots \alpha_n}}$,
set $Y^H:= \sum_{1\leq \alpha\leq n+k}Y^\alpha {\partial \over \partial q^\alpha}$ and
$Y^V:= \sum_{1\leq \alpha_1<\cdots <\alpha_n\leq n+k} Y_{\alpha_1\cdots \alpha_n}
{\partial \over \partial p_{\alpha_1\cdots \alpha_n}}$. Let $X=X_1\wedge \cdots \wedge X_n
\in D^n_m\left( \Lambda^nT^{\star}{\cal N}\right)$ and let us use this decomposition
to each $X_\mu$: then $X$ can be split as $X=\sum_{j=0}^nX_{(j)}$,
where each $X_{(j)}$ is homogeneous of degree
$j$ in the variables $X_\mu^V$ and homogeneous of degree $n-j$ in the
variables $X_\mu^H$.\\

\noindent
Recall that a decomposable $n$-vector $X$
is in $[X]^{\cal H}_m$ if and only if $X\iN \Omega = (-1)^nd{\cal H}$. This equation actually splits as
\begin{equation}\label{3.1.dph}
X_{(0)}\iN \Omega = (-1)^nd_p{\cal H}
\end{equation}
and
\begin{equation}\label{3.1.dqh}
X_{(1)}\iN \Omega = (-1)^nd_q{\cal H}.
\end{equation}
Equation (\ref{3.1.dph}) determines in an unique way $X_{(0)}\in
D^n_q{\cal N}$. The condition $d_p{\cal H}\neq 0$ implies that
necessarily\footnote{Note also that (\ref{3.1.dph}) implies that
$d_p{\cal H}$ must satisfy some compatibility conditions since
$X_{(0)}$ is decomposable.} $X_{(0)}\neq 0$. At this stage we can
choose a family of $n$ linearly independent vectors $X^0_1,\cdots
,X^0_n$ in $T_q{\cal N}$ such that $X^0_1\wedge \cdots \wedge
X_n^0=X_{(0)}$. Thus the forms $dq^\alpha$ are proper on
$[X]^{\cal H}_m$, since their restriction on $X$ are fully
determined by their restriction on the vector subspace spanned by
$X^0_1,\cdots ,X^0_n$. Furthermore the subspace of $T_m{\cal M}$
spanned by $X$ is a graph over the subspace of
$T_q{\cal N}$ spanned by $X_{(0)}$. This proves the part (i) of the Lemma.\\

\noindent Proving (ii) and (iii) requires more work. First
we deduce that there exists a unique family
$(X_1,\cdots ,X_n)$ of vectors in $T_m{\cal M}$ such that $\forall \mu$,
$X_\mu^H=X_\mu^0$ and $X_1\wedge \cdots \wedge X_n=X$.
And Equation (\ref{3.1.dqh}) consists in further underdetermined
conditions on the vertical components $X_{\mu,\alpha_1\cdots \alpha_n}$
of the $X_\mu$'s, namely
\[
\sum_{\mu}\sum_{\alpha_1<\cdots <\alpha_n}C_\beta^{\mu,\alpha_1\cdots \alpha_n}X_{\mu,\alpha_1\cdots \alpha_n}
= - {\partial {\cal H}\over \partial q^\beta},
\]
where
\[
C_\beta^{\mu,\alpha_1\cdots \alpha_n}:= \sum_\nu \delta^{\alpha_\nu}_\beta(-1)^{\mu+\nu}
\Delta^{\alpha_1\cdots \widehat{\alpha_\nu}..\alpha_n}_{1\cdots \widehat{\mu}\cdots n}
\]
and
\[
\Delta^{\alpha_1\cdots \alpha_{n-1}}_{\mu_1\cdots \mu_{n-1}}:= \left|
\begin{array}{ccc}
X^{\alpha_1}_{\mu_1} & \dots & X^{\alpha_1}_{\mu_{n-1}}\\
\vdots & & \vdots \\
X^{\alpha_n}_{\mu_1} & \dots & X^{\alpha_{n-1}}_{\mu_{n-1}}
\end{array}
\right| .
\]
{\em Step2 --- Local coordinates}.
To further understand these relations we choose suitable coordinates $q^\alpha$
in such a way that $d_p{\cal H}_m=dp_{1\cdots n}$ and
\begin{equation}\label{3.1.dph1}
X_\mu^H={\partial \over \partial q^\mu}\quad \hbox{for}\quad \mu =1,...,n,
\end{equation}
so that (\ref{3.1.dph}) is automatically satisfied. In this setting we also have
$$(-1)^nX_{(1)}\iN \Omega = -\sum_\mu X_{\mu,1\cdots n}dq^\mu
- (-1)^n\sum_\mu \sum_{n<\beta}(-1)^\mu X_{\mu,1\cdots \widehat{\mu}\cdots n\beta}dq^\beta,$$
and so (\ref{3.1.dqh}) is equivalent to
\begin{equation}\label{3.1.dqh1}
\left\{ \begin{array}{ccl}
X_{\mu,1\cdots n} & =  & \displaystyle -{\partial {\cal H}\over \partial q^\mu},\;
\hbox{for } 1\leq \mu \leq n\\
 & & \\
\displaystyle
(-1)^n\sum_\mu  (-1)^\mu X_{\mu,1\cdots \widehat{\mu}\cdots n\beta} & = &
\displaystyle - {\partial {\cal H}\over \partial q^\beta},\;
\hbox{for } n+1\leq \beta \leq n+k.
\end{array}\right.
\end{equation}
Let us introduce some notations:
$I:=\{(\alpha_1,\cdots ,\alpha_n)/1\leq \alpha_1<\cdots \leq \alpha_n\leq n+k\}$,
$I^0:=\{(1,\cdots ,n)\}$,
$I^*:=\{(\alpha_1,\cdots ,\alpha_{n-1},\beta)/1\leq \alpha_1<\cdots <\alpha_{n-1}\leq n,
n+1\leq \beta\leq n+k\}$,
$I^{**}:=I\setminus \left( I^0 \cup I^*\right).$
We note also $M_\mu:= \sum_{(\alpha_1,\cdots ,\alpha_n)\in I^*}X_{\mu,\alpha_1\cdots \alpha_n}
\partial ^{\alpha_1\cdots \alpha_n}$,
$R_\mu:= \sum_{(\alpha_1,\cdots ,\alpha_n)\in I^{**}}X_{\mu,\alpha_1\cdots \alpha_n}
\partial ^{\alpha_1\cdots \alpha_n}$ and
$M^\nu_{\mu,\beta}:=(-1)^{n+\nu}X_{\mu,1\cdots \widehat{\nu}\cdots n\beta}$.
Then the set of solutions
of (\ref{3.1.dph}) and (\ref{3.1.dqh}) satisfying (\ref{3.1.dph1}) is
\begin{equation}\label{3.1.xmu}
X_\mu = {\partial \over \partial q^\mu} - {\partial {\cal H}\over \partial q^\mu}
{\partial \over \partial p_{1\cdots n}} + M_\mu +R_\mu,
\end{equation}
where the components of $R_\mu$ are arbitrary, and the coefficients of
$M_\mu$ are only subject to the constraint
\begin{equation}\label{3.1.constraint}
\sum_\mu M^\mu_{\mu,\beta} =
- {\partial {\cal H}\over \partial q^\beta},\quad \hbox{ for } n+1\leq \beta \leq n+k.
\end{equation}

\noindent {\em Step 3 --- The search of all proper 1-forms on $[X]^{\cal H}_m$}.
Now let $a\in T^*_m{\cal M}$ and let us look at
necessary and sufficient conditions for $a$ to be a proper 1-form on $[X]^{\cal H}_m$.
We write
\[
a = \sum_\alpha a_\alpha dq^\alpha + \sum_{\alpha_1<\cdots <\alpha_n}
a^{\alpha_1\cdots \alpha_n}dp_{\alpha_1\cdots \alpha_n}.
\]
Let us write $a^*:=
\left( a^{\alpha_1\cdots \alpha_n}\right)_{(\alpha_1,\cdots ,\alpha_n)\in I^*}$,
$a^{**}:=\left( a^{\alpha_1\cdots \alpha_n}\right)_{(\alpha_1,\cdots ,\alpha_n)\in I^{**}}$
and
\[
\left\langle M_\mu,a^*\right\rangle :=\sum_\nu\sum_{n<\beta}(-1)^{n+\nu}
M^\nu_{\mu,\beta}a^{1\cdots \widehat{\nu}\cdots n\beta},
\]
and
\[
\left\langle R_\mu,a^{**}\right\rangle :=
\sum_{(\alpha_1,\cdots ,\alpha_n)\in I^{**}}X_{\mu,\alpha_1\cdots \alpha_n}
a^{\alpha_1\cdots \alpha_n}.
\]
Using (\ref{3.1.xmu}) we obtain that
\[
a(X_\mu) = a_\mu -
{\partial {\cal H}\over \partial q^\mu}
a^{1\cdots n} +
\left\langle M_\mu,a^*\right\rangle +
\left\langle R_\mu,a^{**}\right\rangle .
\]
\begin{lemm}\label{4.1/2.lemme2}
Condition (\ref{4.1/2.differe}) (resp. (\ref{4.1/2.nul}))
is equivalent to the two following conditions:
\begin{equation}\label{3.1.nul}
a^* =a^{**} = 0
\end{equation}
and
\begin{equation}\label{3.1.nonnul}
\left( a_1 - {\partial {\cal H}\over \partial q^1} a^{1\cdots n},
\; \cdots \; ,
a_n - {\partial {\cal H}\over \partial q^n}
a^{1\cdots n} \right)
\neq 0 \quad \hbox{(resp. }= 0\hbox{)}.
\end{equation}
\end{lemm}
{\em Proof} --- We first look at necessary and sufficient conditions on for $a$
to be a point-slice, i.e.\,to satisfy (\ref{4.1/2.differe}).
Let us denote by $\vec{A}:=\left( a_\mu - {\partial {\cal H}\over \partial q^\mu}
a^{1\cdots n}\right)_\mu$ and
$\vec{M}:=\left( M_\mu\right)_\mu$, $\vec{R}:=\left( R_\mu\right)_\mu$.
We want conditions on $a^{\alpha_1\cdots \alpha_n}$
in order that the image of the affine map $(\vec{M},\vec{R})\longmapsto
\vec{\cal A}(\vec{M},\vec{R}):= \vec{A} +
\langle \vec{M},a^*\rangle + \langle \vec{R},a^{**}\rangle$ does not
contain 0 (assuming that $\vec{M}$ satisfies the constraint (\ref{3.1.constraint})).
We see immediately that if $a^{**}$
would be different from 0, then by choosing $\vec{M}=0$ and $\vec{R}$
suitably, we could have $\vec{\cal A}(\vec{M},\vec{R})=0$. Thus $a^{**}=0$.
Similarly, assume by contradiction that $a^*$ is different from 0. Up
to a change of coordinates, we can assume that
$\left( a^{1\cdots \widehat{\nu}\cdots n(n+1)}\right)_{1\leq \nu\leq n}\neq 0$.
And by another change of coordinates, we can further assume that
$a^{2\cdots n(n+1)} = \lambda \neq 0$ and
$a^{1\cdots \widehat{\nu}\cdots n(n+1)} = 0$, if $\nu\geq 1$.
Then choose $M^\nu_{\mu,\beta}= 0$ if $\beta\geq n+2$, and
\[
\left(\begin{array}{ccccc}
M^1_{1,n+1} & M^1_{2,n+1} & M^1_{3,n+1} & \cdots & M^1_{n,n+1} \\
M^2_{1,n+1} & M^2_{2,n+1} & M^2_{3,n+1} & \cdots & M^2_{n,n+1} \\
\vdots & \vdots & \vdots &  & \vdots \\
M^n_{1,n+1} & M^n_{2,n+1} & M^n_{3,n+1} & \cdots & M^n_{n,n+1}
\end{array}\right) =
\left(\begin{array}{ccccc}
t_1 & t_2 & t_3 & \cdots & t_n\\
0 & s & 0 & \cdots & 0\\
\vdots & \vdots & \vdots &  & \vdots \\
0 & 0 & 0 & \cdots & 0
\end{array}\right),
\]
where $s = - t_1 -\partial {\cal H}/\partial q^{n+1}$. Then we find that
${\cal A}_\mu(\vec{M},\vec{R}) = A_\mu +(-1)^{n+1}\lambda t_\mu$, so that this expression
vanishes for a suitable choice of the $t_\mu$'s. Hence we get a contradiction. Thus
we conclude that $a^*=0$ and $\vec{A}\neq 0$. The analysis of 1-forms which
satisfies (\ref{4.1/2.nul}) is similar: this condition is equivalent to $a^*=0$ and
$\vec{A}=0$.
\bbox
\noindent
{\em Conclusion}. We translate the conclusion of Lemma \ref{4.1/2.lemme2} without
using local coordinates: it gives relation (\ref{4.1/2.base}). \bbox

\subsection{Slices and $r$-regular functions}
\noindent
As an application of the above analysis we can give a characterization of $r$-regular
functions. We first consider the case $r=1$.\\

\noindent
Indeed any smooth function $f:{\cal M}\longrightarrow \Bbb{R}$ is 1-regular if and only
if $\forall m\in {\cal M}$, $df_m$ is a point-slice. Using Lemma
\ref{4.1/2.lemme2} we obtain two conditions on $df_m$:
the condition (\ref{3.1.nul}) can be restated as follows: for all $m\in {\cal M}$
there exists a real number $\lambda(m)$ such that $d_pf_m=\lambda(m)d_p{\cal H}_m$.
Condition (\ref{3.1.nonnul}) is equivalent to: $\exists (\alpha_1,\cdots ,\alpha_n)\in I$,
$\exists 1\leq \mu\leq n$,
\begin{equation}\label{4.2.supercrochet}
\{{\cal H},f\}^{\alpha_1\cdots \alpha_n}_{\alpha_\mu}(m) : =
{\partial {\cal H}\over \partial p_{\alpha_1\cdots \alpha_n}}(m){\partial f\over \partial q^{\alpha_\mu}}(m)
-
{\partial f\over \partial p_{\alpha_1\cdots \alpha_n}}(m){\partial {\cal H}\over \partial q^{\alpha_\mu}}(m)
\neq 0.
\end{equation}
[Alternatively using Lemma \ref{4.1/2.lemme}, $df_m$ is a point-slice
if and only if $\exists \lambda(m)\in \Bbb{R}$, $\exists (a_1,\cdots ,a_{n+k})\in \Bbb{R}^{n+k}$
such that $df_m=\lambda(m)d{\cal H}_m + \sum_{\alpha=1}^{n+k}a_\alpha dq^\alpha$
and $(a_1,\cdots ,a_n)\neq 0$.]
Now we remark
that $d_pf_m=\lambda(m)d_p{\cal H}_m$ everywhere if and only if there exists a
function $\widehat{f}$ of the variables
$(q,h)\in {\cal N}\times \Bbb{R}$ such that $f(q,p)=\widehat{f}(q,{\cal H}(q,p))$.
So we deduce the following.
\begin{theo}\label{4.1/2.theo}
Let ${\cal M}$ be an open subset of $\Lambda^nT^{\star}{\cal N}$ endowed with its
standard multisymplectic form $\Omega$, let ${\cal H}:{\cal M}\longrightarrow \Bbb{R}$
be a smooth Hamiltonian function and let $f:{\cal M}\longrightarrow \Bbb{R}$ be
a smooth function. Assume that $d_p{\cal H}\neq 0$ and $[X]^{\cal H}\neq \emptyset$
everywhere. Then $f$ is 1-regular if and only if there exists a smooth
function $\widehat{f}:{\cal N}\times \Bbb{R}\longrightarrow \Bbb{R}$
such that
\[
f(q,p)=\widehat{f}(q,{\cal H}(q,p)),\quad \forall (q,p)\in {\cal M}
\]
and $\forall m\in {\cal M}$,
\[
\exists (\alpha_1,\cdots ,\alpha_n)\in I, \exists 1\leq \mu\leq n,\quad\quad
\{{\cal H},f\}^{\alpha_1\cdots \alpha_n}_{\alpha_\mu}(m)\neq 0.
\]
\end{theo}
By the same token this result gives sufficient conditions for a hypersurface
defined as the level set $f^{-1}(s):=\{m\in {\cal M}/f(m)=s\}$ of a given function to be
a slice: it suffices that the above condition be true along $f^{-1}(s)$.\\
{\bf Example 8} --- We come back here to critical points $u:{\cal X}\longrightarrow {\cal Y}$
of a Lagrangian functional $l$. We use the notations of Section 2.2 and
denote by $p^*$ the set of coordinates $p^{\mu_1\cdots \mu_j}_{i_1\cdots i_j}$ for $j\geq 1$,
so that ${\cal H}(q,e,p^*) = e+H(q,p^*)$. Let us assume that,
$\forall q\in {\cal N}={\cal X}\times {\cal Y}$, there exists some value $p_0^*$ of $p^*$ such that
$\partial H/\partial p^*(q,p^*_0)=0$. Note that this situation
arises in almost all standard situation (if in particular the Lagrangian $l(x,u,v)$
has a quadratic dependence in $v$). Assume further the hypotheses of Theorem \ref{4.1/2.theo}
and consider a 1-regular function $f\in {\cal C}^\infty({\cal M},\R)$.
We note that $f(q,p) = \widehat{f}(q,{\cal H}(q,p))$ implies that
$\{{\cal H},f\}^{\alpha_1\cdots \alpha_n}_{\alpha_\mu}(q,p) =
{\partial {\cal H}\over \partial p_{\alpha_1\cdots \alpha_n}}(q,p)
{\partial \widehat{f}\over \partial q^{\alpha_\mu}}(q,{\cal H}(q,p))$.
Now for all $(q,h)\in {\cal N}\times \Bbb{R}$, let $p^*_0$ be such that
$\partial H/\partial p^*(q,p^*_0)=0$ and let $e_0:= h -H(q,p_0^*)$. Since
${\partial {\cal H}\over \partial p^*}(q,e_0,p_0^*) = 0$ and ${\partial {\cal H}\over \partial e} = 1$,
condition (\ref{4.2.supercrochet}) at $m = (q,e_0,p_0^*)$ means that $\exists \mu$ with
$1\leq \mu\leq n$ such that ${\partial \widehat{f}\over \partial x^\mu}(q,h) =
{\partial \widehat{f}\over \partial x^\mu}(q,{\cal H}(q,e_0,p_0^*))\neq 0$.
This singles out {\bf space-time coordinates}: they are the functions on ${\cal M}$
needed to build slices.\\

\noindent
We now turn to the case where $1\leq r\leq n$. We consider a map $f=(f^1,\cdots ,f^{r})$
from ${\cal M}$ to $\Bbb{R}^{r}$ and look for necessary and sufficient conditions on $f$
for being $r$-regular. We still assume that $d_p{\cal H}\neq 0$ and $[X]^{\cal H}\neq \emptyset$.
We first analyze the situation locally.
Given a point $m\in {\cal M}$, the property ``$X\in [X]^{\cal H}$ $\Longrightarrow$
$df_{m|X}$ is of rank $r$'' is equivalent to:
\[
\forall (t_1,\cdots ,t_{r})\in \Bbb{R}^{r}\setminus \{0\},\quad
X\in [X]^{\cal H} \Longrightarrow \sum_{i=1}^{r}t_idf^i_{m|X}\neq 0.
\]
Hence by using Lemma \ref{4.1/2.lemme} we deduce that the property
$X\in [X]^{\cal H} \Longrightarrow$ rank $df_{m|X}=r$ is equivalent to
\begin{itemize}
\item $\forall (t_1,\cdots ,t_{r})\in \Bbb{R}^{r}\setminus \{0\}$,
$\exists \lambda(m)\in \Bbb{R}$,
$\sum_{i=1}^{r}t_id_pf^i_m=\lambda(m)d_p{\cal H}_m$.
And then one easily deduce that $\exists \lambda^1(m),\cdots ,\lambda^{r}(m)\in \Bbb{R}$,
such that $\lambda(m)=\sum_{i=1}^{r}t_i\lambda^i(m)$.
\item $\forall (t_1,\cdots ,t_{r})\in \Bbb{R}^{r}\setminus \{0\}$,
$\exists (\alpha_1,\cdots ,\alpha_n)\in I, \exists 1\leq \mu\leq n$,
$\{{\cal H},\sum_{i=1}^{r}t_if^i\}^{\alpha_1\cdots \alpha_n}_{\alpha_\mu}(m)\neq 0$.
\end{itemize}
Now the second condition translate as $\forall (t_1,\cdots ,t_{r})\in \Bbb{R}^{r}\setminus \{0\}$,
$\exists (\alpha_1,\cdots ,\alpha_n)\in I, \exists 1\leq \mu\leq n$,
\[
\sum_{i=1}^{r}t_i{\partial {\cal H}\over \partial p_{\alpha_1\cdots \alpha_n}}
\left( {\partial f^i\over \partial q^{\alpha_\mu}} -\lambda^i
{\partial {\cal H}\over \partial q^{\alpha_\mu}}\right) \neq 0.
\]
This condition can be expressed in terms of minors of size $r$ from the
matrix
$\left(
{\partial f^i\over \partial q^{\alpha_\mu}} -\lambda^i
{\partial {\cal H}\over \partial q^{\alpha_\mu}}\right)_{i,\alpha_\mu}$.
For that purpose let us denote by\\

\noindent
$\displaystyle \{\{ {\cal H},f^1,\cdots ,f^{r}\}\} :=
\sum_{1\leq \alpha_1<\cdots <\alpha_n\leq n+k}\;\sum_{1\leq \mu_1<\cdots <\mu_{r}\leq n}$
$$\left\langle {\partial \over \partial p_{\alpha_1\cdots \alpha_n}}
\wedge {\partial \over \partial q^{\alpha_{\mu_1}}}\wedge \cdots
\wedge {\partial \over \partial q^{\alpha_{\mu_{r}}}},
d{\cal H}\wedge df^1\wedge \cdots \wedge df^{r}\right\rangle
dp_{\alpha_1\cdots \alpha_n}\wedge dq^{\alpha_{\mu_1}}\wedge \cdots  \wedge
dq^{\alpha_{\mu_{r}}}.$$
We deduce the following.
\begin{prop}
Let ${\cal M}$ be an open subset of $\Lambda^nT^{\star}{\cal N}$ endowed with its
standard multisymplectic form $\Omega$, let ${\cal H}:{\cal M}\longrightarrow \Bbb{R}$
be a smooth Hamiltonian function and let $f:{\cal M}\longrightarrow \Bbb{R}^{r}$ be
a smooth function. Let $m\in {\cal M}$ and assume that
$d_p{\cal H}\neq 0$ and $[X]^{\cal H}\neq \emptyset$ everywhere. Then
$X\in [X]^{\cal H}$ $\Longrightarrow$
$df_{m|X}$ is of rank $r$ if and only if
\begin{itemize}
\item
$\exists \lambda^1(m),\cdots ,\lambda^{r}(m)\in \Bbb{R}$,
$\forall 1\leq i\leq r$, $d_pf^i_m=\lambda^i(m)d_p{\cal H}_m$.
\item $\{\{ {\cal H},f^1,\cdots ,f^{r}\}\} (m)\neq 0$.
\end{itemize}
\end{prop}
And we deduce the global result:
\begin{theo}\label{3.1.theo}
Let ${\cal M}$ be an open subset of $\Lambda^nT^{\star}{\cal N}$ endowed with its
standard multisymplectic form $\Omega$, let ${\cal H}:{\cal M}\longrightarrow \Bbb{R}$
be a smooth Hamiltonian function and let $f:{\cal M}\longrightarrow \Bbb{R}^{r}$ be
a smooth function. Assume that $d_p{\cal H}\neq 0$ and $[X]^{\cal H}\neq \emptyset$
everywhere. Then  $f$ is $r$-regular if and only if there exists a smooth
function $\widehat{f}:{\cal N}\times \Bbb{R}\longrightarrow \Bbb{R}^{r}$
such that $f(q,p)=\widehat{f}(q,{\cal H}(q,p))$
and $\forall m\in {\cal M}$, $\{\{ {\cal H},f^1,\cdots ,f^{r}\}\} (m)\neq 0$.
\end{theo}

\subsection{Generalized pseudofibers directions}
We are now able to prove the equivalence in (an open subset of) ${\cal M} = \Lambda^nT^*{\cal N}$
between the two possible definitions of $L^{\cal H}_m$:
either $\left(T_zD^n_q{\cal N}\right)^\perp$ or
\[
\left(T_{[X]^{\cal H}_m}D^n_m{\cal M}\iN \Omega\right)^\perp:=
\{ \xi\in T_{(q,p)}{\cal M}/\ \forall X\in [X]^{\cal H}_{(q,p)},
\forall \delta X\in T_XD^n_{(q,p)}{\cal M},\ \xi\iN \Omega (\delta X) = 0\}
\]
as presented in Sections 2.3 and 3.3. First recall that the Legendre correspondence hypothesis
implies here that $Z_q(p)$ is reduced to a point that we shall denote by $Z(p,q)$.
As a preliminary we prove the following:

\begin{lemm}\label{4.4.lemm2}
Let ${\cal M}$ be an open subset of $\Lambda^nT^{\star}{\cal N}$ and let ${\cal H}$ be
an {\em arbitrary} smooth function from ${\cal M}$ to $\Bbb{R}$, such that $d_p{\cal H}$
never vanishes.
Let $\xi\in L^{\cal H}_{(q,p)}$, then $dq^\alpha(\xi) = 0$, $\forall \alpha$, i.e.
$$\xi = \sum_{\alpha_1<\cdots <\alpha_n}\xi_{\alpha_1\cdots \alpha_n}
{\partial \over \partial p_{\alpha_1\cdots \alpha_n}}.$$
\end{lemm}
{\em Proof} --- We use the results proved in Section 4.1: we know that we can assume
w.l.g.\,that $d_p{\cal H}=dp_{1\cdots n}$. Then any $n$-vector $X\in D^n_{(q,p)}{\cal M}$
such that $(-1)^nX\iN \Omega = d{\cal H}$ can be written $X=X_1\wedge \cdots \wedge X_n$, where
each vector $X_\mu$ is given by (\ref{3.1.xmu}) with the conditions on $M^\nu_{\mu,\beta}$
and $R_\mu$ described in Section 4.1. We construct a solution  $X$ of
$(-1)^nX\iN \Omega = d{\cal H} = \sum_\alpha{\partial {\cal H}\over \partial q^\alpha}
dq^\alpha + dp_{1\cdots n}$ by choosing
\begin{itemize}
\item $R_\mu = 0$, $\forall 1\leq \mu\leq n$
\item $M^\nu_{\mu,\beta} = 0$ if $(\mu,\nu) \neq (1,1)$
\item $M^1_{1,\beta} = - {\partial {\cal H}\over \partial q^\beta}$,
$\forall n+1\leq \beta\leq n+k$
\end{itemize}
in relations (\ref{3.1.xmu}). It corresponds to
$$\left\{ \begin{array}{ccl}
X_1 & = & \displaystyle {\partial \over \partial q^1} - {\partial {\cal H}\over \partial q^1}
{\partial \over \partial p_{1\cdots n}} + (-1)^n \sum_{\beta=n+1}^{n+k}
{\partial {\cal H}\over \partial q^\beta}{\partial \over \partial p_{2\cdots n\beta}}\\
X_\mu & = & \displaystyle {\partial \over \partial q^\mu} -
{\partial {\cal H}\over \partial q^\mu}
{\partial \over \partial p_{1\cdots n}},\quad \hbox{if } 2\leq \mu \leq n.
\end{array}\right.$$
We first choose $\delta X^{(1)}\in T_XD^n_{(q,p)}{\cal M}$ to be
$\delta X^{(1)} := \delta X^{(1)}_1\wedge X_2\wedge \cdots  \wedge X_n$, where
$\delta X^{(1)}_1:={\partial \over \partial p_{1\cdots n}}$. It gives
$$\delta X^{(1)} = {\partial \over \partial p_{1\cdots n}}\wedge {\partial \over \partial q^2}
\wedge \cdots  \wedge {\partial \over \partial q^n}.$$
Now let $\xi\in L^{\cal H}_{(q,p)}$, we must have $\xi\iN \Omega(\delta X^{(1)}) = 0$. But a
computation gives
$$\xi\iN \Omega(\delta X^{(1)}) = (-1)^n\delta X^{(1)}\iN \Omega(\xi)
= -dq^1(\xi),$$
so that $dq^1(\xi) = 0$.\\

\noindent For $n+1\leq \beta\leq n+k$, consider $\delta X^{(\beta)} :=
\delta X^{(\beta)}_1\wedge X_2\wedge \cdots  \wedge X_n\in T_XD^n_{(q,p)}{\cal M}$,
where $\delta X^{(\beta)}_1:={\partial \over \partial p_{2\cdots n\beta}}$. Then
we compute that $\delta X^{(\beta)}\iN \Omega = dq^\beta$. Hence,
by a similar reasoning, the relation
$\xi\iN \Omega(\delta X^{(\beta)}) = 0$ is equivalent to $dq^\beta(\xi) = 0$.\\

\noindent Lastly by considering another solution $X\in D^n_{(q,p)}{\cal M}$ to the Hamilton equation,
where the role of $X_1$ has been exchanged with the role of $X_\mu$, for some
$2\leq \mu\leq n$, we can prove that $dq^\mu(\xi) = 0$, as well. \bbox

\noindent Recall that the tangent space $T_{(q,p)}\left( \Lambda^nT^*{\cal N}\right)$
possesses a canonical ``vertical'' subspace
$\hbox{Ker}d\Pi_{(p,q)}\simeq \Lambda^nT^*_q{\cal N}$:
Lemma \ref{4.4.lemm2} can be rephrased by saying that,
if $d_p{\cal H}\neq 0$ everywhere, then $L^{\cal H}_{(q,p)}$ can be identified
with a vector subspace of this vertical subspace.

\begin{prop}\label{4.4.prop}
Let ${\cal M}$ be an open subset of $\Lambda^nT^{\star}{\cal N}$ and let ${\cal H}$ be a Hamiltonian
function on ${\cal M}$ built from a Lagrangian density $L$ by means of the
Legendre correspondence. Then, through the identification
$\hbox{Ker}d\Pi_{(p,q)}\simeq \Lambda^nT^*_q{\cal N}$,
$\left(T_{[X]^{\cal H}_m}D^n_m{\cal M}\iN \Omega\right)^\perp$ coincides with
$\left(T_{Z(q,p)}D^n_q{\cal N}\right)^\perp$.
\end{prop}
{\em Proof} --- First we remark that the hypotheses imply that
$d_p{\cal H}$ never vanishes (because $d{\cal H}(0,\omega)=1$).
Let $\xi\in \left(T_{[X]^{\cal H}_m}D^n_m{\cal M}\iN
\Omega\right)^\perp$, using the preceding remark we can associate
a $n$-form $\pi \in \Lambda^nT^{\star}_q{\cal N}$ to $\xi$ with
coordinates $\pi_{\alpha_1\cdots \alpha_n} = \xi_{\alpha_1\cdots
\alpha_n}$, simply by the relation $\pi = \xi\iN \Omega$. Now let
us look at the condition:
\begin{equation}\label{4.4.implication}
\forall X\in [X]^{\cal H}_{(q,p)},\quad \forall \delta X\in T_XD^n_{(q,p)}{\cal M},
\quad  \xi\iN \Omega(\delta X)=0.
\end{equation}
By the analysis of section 4.1 we know that the ``horizontal'' part $X_{(0)}$ of $X$
is fully determined by ${\cal H}$: it is actually $X_{(0)} = Z(q,p)$. Now take
any $\delta X\in T_XD^n_{(q,p)}{\cal M}$ and split it into its horizontal part
$\delta z\in T_{Z(q,p)}D^n_q{\cal N}$ and a vertical part $\delta X^V$. We remark that
\begin{itemize}
\item $\delta z\in T_{Z(q,p)}D^n_q{\cal N}$
\item $\xi\iN \Omega(\delta X) = \pi(\delta X) = \pi(\delta z)$.
\end{itemize}
Hence (\ref{4.4.implication}) means that $\pi\in \left( T_{Z(q,p)}D^n_q{\cal N}\right)^\perp$.
So the result follows. \bbox

\end{document}